%% file: Keen_1009.tex
\documentclass{emulateapj}

\usepackage{graphicx} \usepackage{natbib}  \usepackage{amsmath} \usepackage{color}
\shorttitle{Pair Fraction Evolution at $Z<1$}

\begin{document}

\title{Evolution of the Major Merger Galaxy Pair Fraction at $z < 1$}
\author{R.~C.~Keenan\altaffilmark{1}, S. Foucaud\altaffilmark{2}, R. De Propris\altaffilmark{3}, B.~C.~Hsieh\altaffilmark{1}, L. Lin\altaffilmark{1}, R.~C.~Y.~Chou\altaffilmark{1}, S. Huang\altaffilmark{1}, J.~H. Lin\altaffilmark{1}, K.~H. Chang\altaffilmark{1}}

\altaffiltext{1}{Academia Sinica Institute for Astronomy and Astrophysics, Taipei, Taiwan} 
\altaffiltext{2}{Shanghai Jiao Tong University, Shanghai, China}
\altaffiltext{3}{Finnish Centre for Astronomy with ESO (FINCA), University of Turku, Finland}

\begin{abstract}
We present a study of the largest available sample of near-infrared selected (i.e., stellar mass selected) dynamically close pairs of galaxies at low redshifts ($z<0.3$).  We combine this sample with new estimates of the major-merger pair fraction for stellar mass selected galaxies at $z<0.8$, from the Red Sequence Cluster Survey (RCS1). We construct our low-redshift $K-$band selected sample using photometry from the UKIRT Infrared Deep Sky Survey (UKIDSS) and the Two Micron All Sky Survey (2MASS) in the $K-$band ($\sim 2.2~\mu$m).  Combined with all available spectroscopy, our $K-$band selected sample contains $\sim 250,000$ galaxies and is $> 90\%$ spectroscopically complete.   The depth and large volume of this sample allow us to investigate the low-redshift pair fraction and merger rate of galaxies over a wide range in $K-$band luminosity.   We find the major-merger pair fraction to be flat at $\sim 2\%$ as a function of $K-$band luminosity for galaxies in the range $10^8 - 10^{12} L_{\odot}$, in contrast to recent results from studies in the local group that find a substantially higher low-mass pair fraction.   
This low-redshift major-merger pair fraction is $\sim 40-50\%$ higher than previous estimates drawn from $K-$band samples, which were based on 2MASS photometry alone.    Combining with the RCS1 sample we find a much flatter evolution ($m = 0.7 \pm 0.1$), in the relation $f_{\rm{pair}} \propto (1+z)^m$, than indicated in many previous studies.  These results indicate that a typical $L\sim L^*$ galaxy has undergone $\sim 0.2-0.8$ major mergers since $z=1$ (depending on the assumptions of merger timescale and percentage of pairs that actually merge).  
  
\end{abstract} 
\maketitle

\section{Introduction}
The galaxy major-merger rate and its evolution are important quantities for theories of galaxy formation.  In hierarchical cold dark matter models that include a cosmological constant ($\Lambda$CDM models), galaxies are expected to accrete most of their stellar mass via mergers, with at least $50\%$ of the total stellar mass growth occurring at $z<1$ \citep{Delu06, Delu07}.  Major mergers should have a profound influence on galaxy properties such as morphology, star formation rate, and nuclear activity, among others (e.g., \citealt{Toom72, Heck86, Sand88}).  In particular, mergers and interactions are implicated in the transformation of field galaxies, which are typically disk-dominated, blue, and star-forming,  into cluster galaxies, which are typically bulge dominated, red, and quiescent \citep{Bell04, Fabe07, Brow07, Heid09}. 

The evolution of the major-merger rate is commonly characterized via a determination of the fraction of galaxies in bound pairs as a function of redshift, which is parameterized as $f_{\rm{pair}} \propto (1+z)^m$.  Studies seeking to understand the evolution of the galaxy merger rate have arrived at values for the exponent spanning a wide range ($0<m<5$, e.g., \citealt{Lin04, Lin08, Kart07, Dera09}), and thus, the redshift evolution of the merger rate remains largely unconstrained.  In recent review, \citet{Cons14} gives a detailed overview of the various discrepancies between studies and their likely origins.  Some of these discrepancies are due to different selection methods or different criteria for identifying merger candidates, but sample variance, redshift incompleteness, and other biases may also be playing a role.  Simulations suggest that the dark matter halo merger rate should evolve with an exponent of $m \approx 2-3$ \citep{Gott01, Fakh08, Gene09, Fakh10}.  However, translating the halo merger rate into the merger rate observed for galaxies, given particular selection criteria and methods, is not straightforward \citep{Lotz11}.  

In general, the two ways of identifying mergers in a sample of galaxies are either to select physically close pairs of galaxies, or galaxies that appear morphologically disturbed.  While a pair selection tends to identify early stage mergers, a  selection by morphology identifies late stage mergers or post-merger galaxies.  Here, we elect to focus on close pairs of galaxies that can be classified as major mergers (defined here as a maximum luminosity ratio of $L_{\rm{primary}}/L_{\rm{secondary}} < 10^{0.4}$), given that the pair fraction of such systems is one of the best proxies for the halo merger rate measured in CDM models \citep{Gene09}.  We also compare with the catalog of merging galaxies selected morphologically in the Galaxy Zoo project by \citet{Darg10a} to quantify how very close projected pairs may be missed in our selection.

Dynamically close pairs of galaxies are those having a projected separation on the sky and line of sight velocity difference such that they have a significant probability of being a bound system.   Typical selection criteria for such pairs are a projected separation of $< 20~h^{-1}$~kpc and velocity difference of $< 500~$km~s$^{-1}$ \citep{Patt97, Patt00, Patt02, Xu04, Depr05,Depr07, Depr10,Kart07, Patt08, Xu12}.  Simulations have shown that the vast majority of pairs meeting these criteria will merge on timescales of $< 1$~Gyr \citep{Boyl08, Kitz08,Jian12}.  Other groups have used different selection criteria, including wider projected separation or larger velocity difference, as well as including consideration of morphology or the two-point correlation function \citep{Frie88, Xu91, Lefe00, Lin04,Lin07,Lin08,Lin10,Bell06, Li08, Domi09, Roba10, Darg10a,Darg10b,Elli08, Elli11, Elli13b,Elli13a, Scud12,Patt11,Patt13, Saty14}.

Here we identify dynamically close pairs from flux limited samples selected in the $K-$band ($2.2~\mu$m) using photometry from the UKIRT Infrared Deep Sky Large Area Survey (UKIDSS-LAS, \citealt{Lawr07})  and the Two Micron All Sky Survey Extended Source Catalog (2MASS-XSC; \citealt{Skru06}).  
Such a sample is effectively stellar mass selected, because near-infrared (NIR) luminosity is well correlated with stellar mass \citep{Gava96, Dejo96, Bell01, Bell03, Kirb08}.  In this study, we consider various quantities, such as the pair fraction and merger rate, in terms of $K-$band luminosity rather than stellar mass, but the standard practice to convert NIR luminosity to stellar mass simply involves applying a single conversion factor for all galaxies based on the assumption of a particular stellar initial mass function and star formation history (e.g., \citealt{Domi09, Xu12}).  

The combination of NIR photometry with all publicly available spectroscopy allows us to construct a stellar mass selected sample with both the depth and area on the sky to measure the major-merger pair fraction in galaxies over a wide range in stellar mass.  We  combine this sample with a stellar mass selected sample from the Red Sequence Cluster Survey (RCS1, \citealt{Glad05}) to investigate the evolution of the pair fraction of $L\sim L^*~(10^{11}~L_{\odot})$ galaxies at $z<0.8$. 

On the faint end of our NIR selected sample, we can probe down to luminosities of a few times $10^8~L_{\odot}$, or roughly half the NIR luminosity of the Large Magellenic Cloud, out to distances of $\sim 50~$Mpc.   This is of particular interest because recently, \citet{Fatt13} showed that $\sim 30\%$ of dwarf galaxies (stellar masses $M < 10^{9.5}~M_{\odot}$) in the local group reside in close pairs, in which the galaxies are of comparable luminosity (less than three magnitudes difference).  They point out that the expectation from $\Lambda$CDM models is that galaxy formation efficiency should be reduced dramatically with decreasing halo mass, such that on dwarf galaxy scales, physical pairs of similar luminosity should be rare.   

They go on to show that simulations predict the local group dwarf pair fraction should be $\sim 4\%$ (not exceeding $12\%$ in $>1000$ realizations of the simulations).     The sample we consider here contains $\sim 1000$ galaxies in the mass range $10^8 -10^{9.5} M_{\odot}$, which allows us to investigate the incidence of pairs among dwarf galaxies in the local universe.

We describe the sample selection in Section~\ref{catgen}, the identification of galaxy pairs in Section~\ref{pairs}, the calculation of the pair fraction and merger rates in Section~\ref{pairfrac}, and we summarize in Section~\ref{summary}. 
All magnitudes given in this paper are in the AB magnitude system ($K_{\rm{AB}} = K_{\rm{Vega}} + 1.9$, $K_{\rm{s,AB}} = K_{\rm{s,Vega}} + 1.86$), where $m_{\rm{AB}} = 23.9-2.5~\rm{log}_{10}(\emph{f}_\nu)$ with $f_\nu$ in units of $\mu$Jy.  We assume a cosmology of $\Omega_M = 0.3, \Omega_{\Lambda} = 0.7,~$and$~H_0 = 100h$ km~s$^{-1}$~Mpc$^{-1}$, with $h=0.7$ in our conversion of redshifts to distances.  We retain the ``little $h$" in our projected separation criterion ($5 < r_{\rm{sep}} < 20~h^{-1}$ kpc) for ease of comparison with many previous studies.

\section{Sample Selection}
\label{catgen}

\begin{figure*}
\begin{center}
\includegraphics[width=170mm]{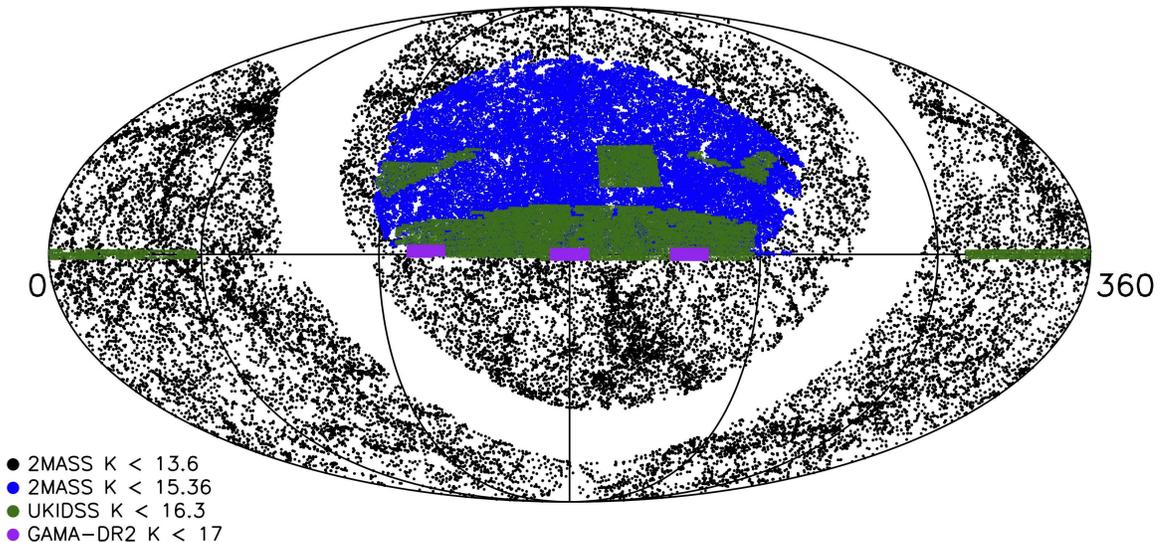}
\caption{\label{cover} The source distribution on the sky for the low-redshift samples considered in this study.  The 2MASS samples are shown in black and blue, the UKIDSS sample in green, and the GAMA sample in purple.}
\end{center}
\end{figure*} 

The low-redshift samples we study here are drawn from publicly available data and include NIR photometry from UKIDSS  and  2MASS combined with redshifts from a variety of optical spectroscopic surveys.  The coverage on the sky of each of the subsamples is shown in Figure~\ref{cover}.

\subsection{UKIDSS Sample}

The UKIDSS-LAS contains NIR photometry in the $Y, J, H,$ and $K-$bands over $\sim 3000$ square degrees on the sky to a $K-$band depth of $K_{\rm{AB}} \sim 20.1$.   UKIDSS uses the UKIRT Wide Field Camera (WFCAM, \citealt{Casa07}).  The photometric system is described in \citet{Hewe06}, the calibration is described in \citet{Hodg09}, and the pipeline processing and archive are described in \citet{Hamb08}.  Much of the UKIDSS-LAS area coincides with the footprint of the Sloan Digital Sky Survey (SDSS, \citealt{York00}), where spectroscopy of bright galaxies is highly complete.  

We generated our wide-area UKIDSS sample by combining photometry from data release 9 (DR9) of the UKIDSS-LAS with redshifts from data release 10 (DR10) of the SDSS, the Two-Degree Field Galaxy Redshift Survey (2DFGRS; \citealt{Coll01}), and other publicly available data.  We selected only objects from areas on the sky that had been imaged in all four UKIDSS bandpasses (to aid in star/galaxy separation), and where a counterpart existed within a radius of $2\arcsec$ in the SDSS catalogs.  This resulting selection of galaxies is $>90\%$ spectroscopically complete to $K_{\rm{AB}}= 16.3$ over $\sim 2000$ square degrees within the SDSS footprint.  Our methods for sample selection, star-galaxy separation, etc., are identical to those described in \citet{Keen13}, except that we have updated to the UKIDSS-DR9 and SDSS-DR10.  
\pagebreak
\subsection{GAMA Sample}
The Galaxy And Mass Assembly Survey (GAMA; \citealt{Driv09, Driv11}) team have recently made their second data release (DR2; J. Liske et al., in preparation), which includes spectroscopy and independent $K-$band photometry performed on UKIDSS imaging \citep{Hill11} of sources located in three equatorial fields totaling $144$~deg$^2$.   GAMA targets for spectroscopy are selected in the $R-$band ($R<19.4$), which provides for high spectroscopic completeness ($\sim 95\%$) in the $K-$band to $K_{\rm{AB}}= 17$. 

\subsection{2MASS Sample}

\citet{Bili14} recently published a redshift catalog (2MPZ) for 2MASS galaxies, which contains spectroscopic redshifts, where available, as well as photometric redshifts for all galaxies to a depth of $K_{\rm{AB}} = 15.76$.  Here we use only the  spectroscopic redshift sample of the 2MPZ, which contains redshifts from the Two Micron Redshift Survey (2MRS; \citealt{Huch05, Erdo06}), the Six-Degree Field Galaxy Redshift Survey (6DFGRS; \citealt{Jone09}), the 2DFGRS, the SDSS, and other publicly available data.  The 2MPZ catalog is $\sim 98\%$ spectroscopically complete down to $K_{\rm{AB}}=13.6$ over the entire extragalactic sky, and $\sim 94\%$ complete down to $K_{\rm{AB}}= 15.36$ over the main footprint of the SDSS.  We update the 2MPZ for our study to include redshifts from the SDSS-DR10 ($\sim 6000$ additional redshifts). 

\subsection{Photometry}
\label{phot}
This study makes a comparison of low-redshift pair fractions and merger rates derived from 3 different sources of photometry (2MASS, UKIDSS, and GAMA).   However, our results do not depend sensitively on the initial photometry catalogs used.

For the main UKIDSS sample, we use $K-$band Petrosian aperture magnitudes.  The UKIDSS pipeline imposes an upper limit on the Petrosian aperture radius of $6\arcsec$, which implies a circular aperture radius of $12\arcsec$.  This results in the flux being systematically underestimated for $\sim 10\%$ of the galaxies in our sample which appear large on the sky.  

In \citet{Keen13}, we describe a method for recovering the light lost to this Petrosian aperture ``clipping", which involves fitting and extrapolating S\'{e}rsic profiles to the light curves derived from a range of circular aperture ($1-12\arcsec$) measurements.  There, we demonstrate that this method provides a satisfactory lost-light correction and we employ the same method in this study to correct Petrosian aperture photometry from UKIDSS.  We note, however, that this correction does not change the results or conclusions presented here.  

\citet{Hill11} provided a reanalysis of UKIDSS photometry for sources in the GAMA fields.  In the analysis of the GAMA data presented here we use these updated $K-$band Petrosian aperture magnitudes.   In principle, these should be better than UKIDSS Petrosian aperture magnitudes because they do not suffer from the aperture clipping issue described above.  However, the GAMA-DR2 catalogs provide $K-$band photometry for Petrosian apertures defined in the $r-$band, which may present some bias.

 Between UKIDSS and GAMA Petrosian aperture magnitudes, there is $\sim 0.1$ mag rms scatter and GAMA magnitudes are systematically brighter by $\sim 0.03$ mag.  We ran all of our analysis on the GAMA fields presented below using both UKIDSS and GAMA input photometry catalogs and found no significant difference in our results that depended on which photometry we used.

2MASS used a $K_s$ filter, which features a slightly shorter central wavelength ($2.12~\mu$m) than the UKIDSS $K$ filter ($2.2~\mu$m).  The magnitudes we use in the 2MASS analysis are the $20$ mag arcsec$^{-2}$ circular isophotal magnitudes ($k\_m\_k20fe$, \citealt{Jarr00}).  A comparison between these and UKIDSS Petrosian aperture magnitudes shows that UKIDSS runs, on average, $\sim 0.03$ mag brighter than 2MASS, and with a relatively large scatter of $\sim 0.2$ mag between the two catalogs.  \citet{Domi09} demonstrate that 2MASS photometry becomes problematic for objects which are close in projection on the sky (blended).  We address this issue below in Section~\ref{blendcorr}.

While a direct comparison between 2MASS and UKIDSS photometry could be problematic due to the issues mentioned above, the purpose of the 2MASS photometry in this study is to rerun some of the analyses presented in previous studies as a baseline check of our methods.  The main results of this study are derived from UKIDSS photometry, so we make no further efforts to reconcile the differences between these catalogs.

\subsection{Spectroscopy and Completeness}
\begin{figure}
\begin{center}
\includegraphics[width=90mm]{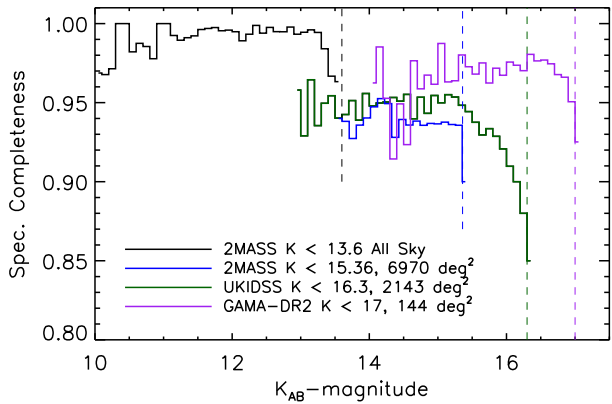}
\caption{\label{complete} Spectroscopic completeness as a function of $K-$band apparent magnitude.  The 2MASS samples are shown in black and blue, the UKIDSS sample in green, and the GAMA sample in purple.  The dashed vertical lines of the same colors indicate the magnitude limit of each sample.}
\end{center}
\end{figure}

This study relies most heavily on spectroscopy from the SDSS, 2DFGRS, and GAMA surveys.  The SDSS spectrograph and the 2DF + AAOmega spectrograph (used for the 2DFGRS and GAMA) feature similar resolution and wavelength coverage in the configurations used for the surveys ($\sim 3800 - 9000~$\AA,~~$R\sim 1800$).  The SDSS team reports a redshift accuracy of $30~$km~s$^{-1}$ \citep{Stou02}, while the 2DFGRS and GAMA surveys report redshift accuracies of $85~$km~s$^{-1}$ \citep{Coll01} and $65 ~$km~s$^{-1}$ \citep{Driv11}, respectively.  

We made a comparison between SDSS and 2DF redshifts for $\sim 10,000$ galaxies in the equatorial overlap region between these surveys.  We found an rms difference of $\sim 100~$km~s$^{-1}$ in the redshifts of galaxies that had been measured by both surveys, and $> 99\%$ agreement to within $\pm 300$ ~km~s$^{-1}$.  In our analysis, where there is overlap we elect to use the SDSS redshifts, given the better stated redshift precision.  

We supplement our redshift catalogs with other publicly available spectroscopic data from the NASA Extragalactic Database (NED\footnote{http://ned.ipac.caltech.edu/}).  However, these extra redshifts make up less than $10\%$ of our sample and we expect that the redshift accuracy of the entire sample should remain better than $\pm 100~$km~s$^{-1}$.  We show below in our analysis that the majority of galaxy pairs in our sample have velocity differences of $< 300~$km~s$^{-1}$, such that our selection criterion of $\Delta V < 500~$km~s$^{-1}$ should be minimally affected by redshift uncertainty.

In Figure~\ref{complete}, we show the spectroscopic completeness ($N_{\rm{specz}}/N_{\rm{total}}$) as a function of  $K-$band apparent magnitude for the various  subsamples of galaxies.  The 2MASS samples are shown in black and blue, the UKIDSS sample in green, and the GAMA sample in purple.  The rough area on the sky of each sample is listed in the plot, and vertical dashed lines indicate the various magnitude limits.  

\subsection{Redshift and Luminosity Distributions}

\begin{figure}
\begin{center}
\includegraphics[width=90mm]{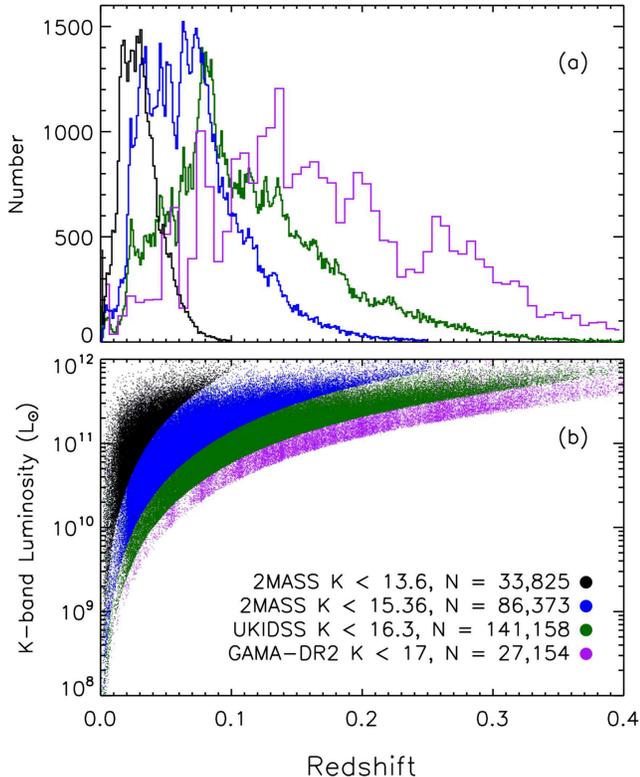}
\caption{\label{zvnsm} (a) The redshift distribution of the low-redshift samples included in this study.  The 2MASS samples are shown in black and blue, the UKIDSS sample in green, and the GAMA sample in purple.  The histograms have been binned such that they are scaled to roughly the same vertical scale in the plot. (b) $K-$band luminosity as a function of redshift for the same samples displayed in (a).  The number of galaxies in each sample is also listed in the plot.}
\end{center}
\end{figure}

The redshift distributions of each of the aforementioned samples is shown in Figure~\ref{zvnsm}a.  The 2MASS samples are shown in black and blue, the UKIDSS sample in green, and the GAMA sample in purple.  The  histograms have been binned such that they appear on roughly the same vertical scale.  The actual number of galaxies in each sample is listed in the lower panel of the plot.

To calculate absolute magnitudes we modified the observed apparent magnitude by a distance modulus $(DM)$, a $K-$correction $K(z)$, and an evolution correction $E(z)$, such that the absolute magnitude of a given galaxy is described by $M=m-DM(z)-K(z)+E(z)$.

In the NIR at low redshifts, $K-$corrections are small and nearly galaxy type independent  \citep{Mann01}.  \citet{Chil10} show, using SDSS and UKIDSS data,  that at low redshifts ($z<0.5$), accurate $K-$corrections can be calculated using inputs of only redshift and one observed color.  They compared with more rigorous methods of spectral energy distribution fitting \citep{Blan07, Fioc97}, and found the errors associated with the $K-$corrections using their simpler method should be $<0.1$~mag.   They have provided a $K-$correction calculator\footnote{http://kcor.sai.msu.ru/}, which we used to estimate the $K-$corrections for our sample.   
 
Evolution of the rest-frame intensity of the NIR light from galaxies is expected to be substantially weaker than for optical bandpasses \citep{Blan03}.  However, we apply an evolution correction to compare galaxy pairs in different redshift bins at their expected $z=0$ luminosities.  The standard evolution correction takes the form $E(z) = Qz$, with $Q$ being a positive constant.  \citet{Blan03} have shown that, for the $K-$band, $Q=1$ fits the expectation from  stellar population synthesis models.  Thus, here we assume a value of $Q=1$, such that $E(z) = z$.

We assume a value for the solar luminosity in the $K-$band of $M_{\odot,\rm{K}}=5.19$, appropriate for the UKIDSS $K$ filter \citep{Hill10}.  While this is not strictly appropriate for the 2MASS $K_s$ filter, we use the same value for consistency, and, as noted in Section~\ref{phot}, the main results of this study are drawn from the UKIDSS sample.  In Figure~\ref{zvnsm}b, we show the distribution of $K-$band luminosities as a function of redshift for the various samples in this study.  The color coding is the same as for the top panel of this figure.
 
\section{Selecting Dynamically Close Pairs of Galaxies}
\label{pairs}

\begin{figure}
\begin{center}
\includegraphics[width=90mm]{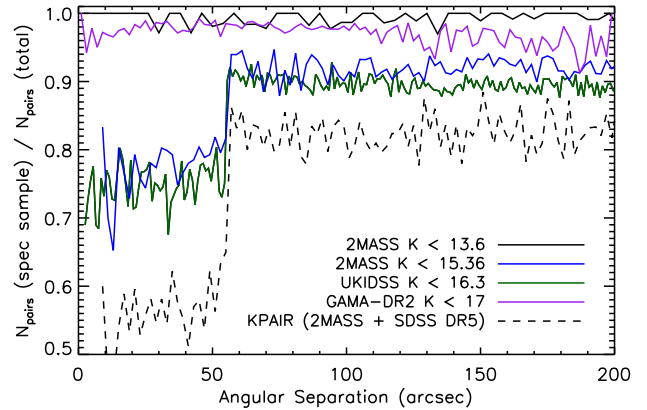}
\caption{\label{pairbias} The ratio of the number of pairs of galaxies, where both galaxies have a spectroscopic redshift, to the number of pairs in the entire sample (regardless of redshift information) as a function of angular separation.  This ratio is then the pair completeness as a function of the angular separation of pairs on the sky.  Fiber collision issues in the SDSS cause the drop in completeness below $55\arcsec$ in the deeper 2MASS sample and the UKIDSS sample.  We also include the same statistic for the KPAIR sample (2MASS + SDSS-DR5, \citealt{Domi09, Xu12}) to demonstrate the improvement in spectroscopic completeness for close pairs in our study compared to previous low-redshift NIR-selected samples. }
\end{center}
\end{figure}

We select dynamically close pairs from the samples described above by requiring that pair galaxies have a projected separation of $5-20~h^{-1}$~kpc ($h=0.7$) and a velocity difference of $\Delta V < 500$~km~s$^{-1}$.  The lower limit of $5~h^{-1}$~kpc is chosen to avoid confusion due to blended pairs, though this lower limit is only expected to exclude $\sim 5\%$ of pairs \citep{Patt97, Patt00}.  We choose the upper limit of $20~h^{-1}~$kpc and $\Delta V < 500$~km~s$^{-1}$~because at least half of such pairs show physical signs of interactions  \citep{Patt00}, and simulations have shown that the vast majority of these pairs will merge on timescales of $\sim 1$~Gyr \citep{Kitz08}.  Furthermore, these selection criteria match those of several previous studies, which facilitates a direct comparison with other results.

\subsection{Spectroscopic Completeness for Pairs}
In any sample that is not 100\% spectroscopically complete, some close pairs will be missed because one or both of the galaxies lack redshift information.  Thus, in principle, such a sample will contain pairs where both galaxies have a redshift, pairs where one galaxy has a redshift, and those where neither have redshifts.  This issue is compounded by the fact that, in general, there exists a bias against measuring redshifts for objects that are close in projection on the sky due to fiber (or slit) ``collisions".

In Figure~\ref{pairbias}, we show the ratio of the number of double-redshift galaxy pairs in each of the spectroscopic samples, to the total number of pairs (regardless of redshift information), as a function of angular separation.  This ratio is then the spectroscopic completeness for galaxy pairs as a function of angular separation on the sky.  We note that for pairs in the all-sky 2MASS sample and the GAMA-DR2 sample, the pair completeness is flat as a function of angular separation.   In the 2MASS sample this is likely due to the fact that these relatively bright galaxy pairs are at wider separation, so fiber collisions are less of a problem, as well as the fact that multiple overlapping redshift surveys were combined in this sample.  In the GAMA-DR2 sample, multiple visits to each of the fields to ensure maximum completeness has alleviated the fiber collision issue.  

However, in both the 2MASS and UKIDSS samples within the SDSS footprint we measure a drop in completeness for pairs at separations of less than $55\arcsec$, where fiber collisions become a problem in the SDSS.   At separations greater than $55\arcsec$ the completeness for pairs is equal to the square of the overall completeness of the samples.  

\begin{figure}
\begin{center}
\includegraphics[width=90mm]{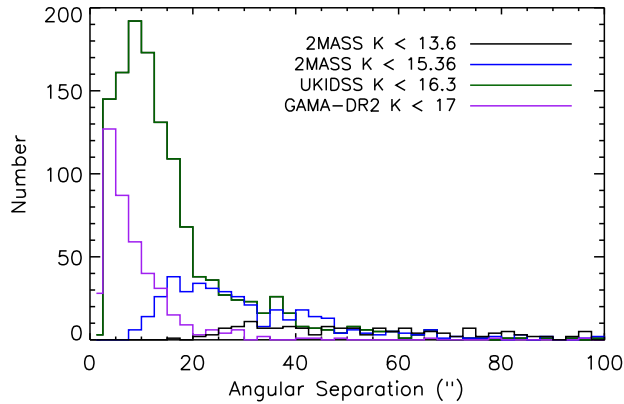}
\caption{\label{asep} Angular separation of pairs on the sky for the various samples.}
\end{center}
\end{figure}

In Figure~\ref{asep}, we show the angular separation for pairs in the various samples.  We note that the vast majority of pairs lie at angular separations of $<55\arcsec$, such that in the 2MASS ($K<15.36$) and UKIDSS samples, there will be a significant number of ``single-redshift pairs", where one galaxy lacks spectroscopy.  

However, in Figure~\ref{pairbias}, we also show completeness for pairs as a function of angular separation for the 2MASS + SDSS-DR5 sample used to generate the ``KPAIR" sample of \citet{Domi09} and \citet{Xu12}.  We show this to highlight the significant improvement, both in overall completeness $(+10\%)$ and for pairs at $<55\arcsec$ separations ($+20\%$) in the UKIDSS (or 2MASS) + SDSS-DR10 (+2DFGRS, +GAMA, etc.) in the samples considered here.  

\subsection{Accounting for Single Redshift Pairs}
\label{singlepair}
In principle, the probability of any single-redshift pair being a physical pair is a function of angular separation, redshift, and the apparent magnitude difference between the pair galaxies.  Here, we wish to consider the pair fraction for major mergers, so we calculate the probability that any given pair of galaxies having less than one magnitude difference in brightness is a physical pair, given its redshift and angular separation.
  
\begin{figure}
\begin{center}
\includegraphics[width=90mm]{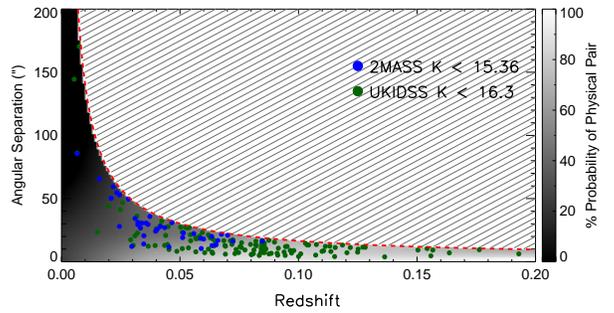}
\caption{\label{probmap} A map of the probability (denoted by the grayscale on the right), of a single-redshift major-merger candidate pair being a physical pair.    Blue points show the distribution in redshift and angular separation for single-redshift major-merger candidate pairs in the 2MASS sample ($K<15.36$) and green points show the same for the UKIDSS sample.  This map was generated using the spectroscopic sample of galaxies by comparing the number of physical pairs to projected pairs both as a function of redshift and angular separation.  The red dashed line shows the selection criterion of pairs being at projected separations of $<20~h^{-1}$ kpc.}
\end{center}
\end{figure}

In Figure~\ref{probmap}, we present the probability of pairs being physical pairs as a function of redshift and angular separation on the sky.   The grayscale denotes the probability of a given galaxy pair being a physical pair as a function of position in this plane.  

We calculated this probability considering only the galaxies in the 2MASS ($K<15.36$) and UKIDSS samples which have redshifts (i.e., only double-redshift pairs).  In these spectroscopic samples, we then counted physical pairs at a given angular separation and redshift (those meeting all of our pair selection criteria)  and divided by the total number of double-redshift pairs at that angular separation and redshift with no velocity constraint.   This exercise yielded the grid of probabilities, as a function of redshift and angular separation, shown in Figure~\ref{probmap}.

The dashed red line in Figure~\ref{probmap} denotes the selection criterion of a projected separation $<20~h^{-1}$ kpc.  Blue points show the location of 2MASS ($K<15.36$) single-redshift pairs and green points show UKIDSS single-redshift pairs.  Given the location of single-redshift pairs in this plane we estimate a probability that any given pair is physical, such that they may be included appropriately in the pair fraction.

This figure demonstrates that the probability of any two galaxies being in a physical pair is a strong function of redshift and angular separation.  While the probability of two galaxies on the low-redshift end of the sample being in a physical pair varies from roughly $0-50\%$, beyond $z\sim 0.1$, pairs at projected separations of $<20~h^{-1}~$kpc are nearly all physical pairs.  The total contribution of single-redshift pairs to the pair fractions calculated below is $\sim 15-20\%$.

\subsection{Correcting for Blending in the 2MASS samples}
\label{blendcorr}

\begin{figure*}
\begin{center}
\includegraphics[width=150mm]{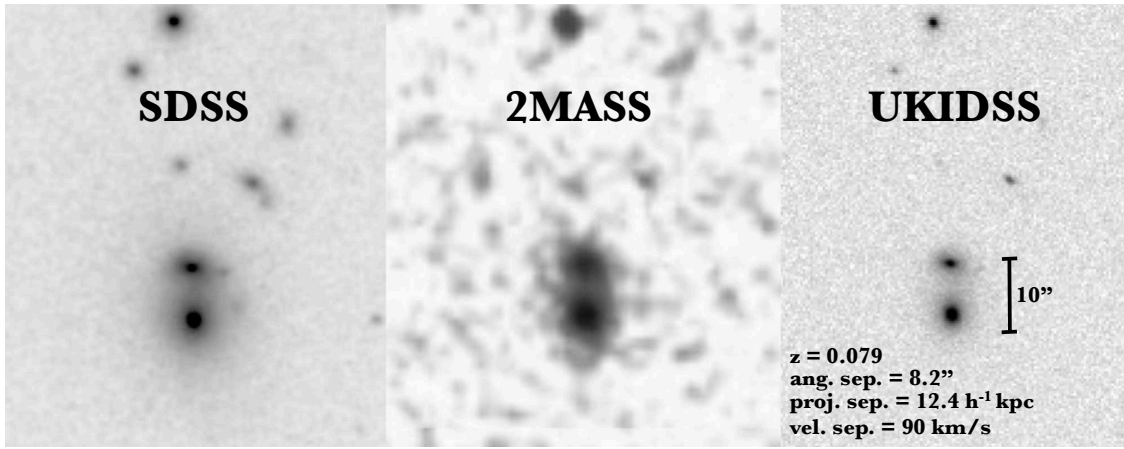}
\caption{\label{blending} A typical major-merger pair that is catalogued as a single object in 2MASS due to blending and is not included as a Galaxy Zoo merger because it does not appear to be a disturbed system.  26 such major-merger pairs exist in the overlap region between 2MASS and UKIDSS, where the primary and secondary meet all the selection criteria for major mergers in the 2MASS $K<15.36$ sample.  UKIDSS only overlaps $\sim 30\%$ of the 2MASS area, suggesting the total number of pairs missed in 2MASS due to this effect is $\sim 90$.}
\end{center}
\end{figure*}

In their consideration of pairs selected from 2MASS + SDSS, \citet{Domi09} note that galaxies separated by less than $10\arcsec$ are treated as single objects in the 2MASS photometric pipeline, resulting in physical pairs at close angular separations being missed due to blending.  To correct for this effect, they visually inspected the 8837 galaxies in their primary sample that have redshifts of $z>0.034$, where objects meeting their selection criteria may be blended.  

Upon inspection, they found 126 candidate pairs and concluded that 51 of these pairs meet their selection criteria.    Through follow-up spectroscopy and other cuts, they exclude roughly half of these additional pairs from their final sample.  However, with a final sample of just 170 pairs, those represented by single objects in the 2MASS catalogs make up $\sim 15\%$ of the total, such that this is one of the most significant corrections made to the raw pair counts.

To investigate the effects of blending in our 2MASS + SDSS sample, we first employed the Galaxy Zoo (GZ) mergers catalogue of \citet{Darg10a}.  The GZ project is described in detail in \citet{Lint08}, and consisted of employing the efforts of $\sim 140,000$ volunteers in the visual classification of $\sim 900,000$ galaxies in the SDSS.  One of the products of this effort was the catalogue of interacting galaxies described in \citet{Darg10a}.  Here we use this GZ mergers catalogue to identify pairs of interacting galaxies that were missed in our initial selection due to blending in 2MASS.  

Volunteers that participated in the GZ project were simply offered the option to classify a system as a ``merger" based on their visual inspection of an image.  A ``weighted-merger-vote fraction, $f_m$", which is a sort of probability that any given object is actually a merger, was then defined for each object based on these results (see \citealt{Darg10a} for more details on $f_m$).  Based on further investigations of morphologies versus $f_m$, the GZ team chose a cutoff in $f_m$ to define their mergers catalogue.  Thus, the GZ selection is by morphology only, and is quite different from our pair selection.  However, this ``by eye" classification of merging systems has been demonstrated to be a reliable measure of galaxy interaction \citep{Darg10a}, and allows us to quantify what fraction of merging systems are missed due to blending in 2MASS.

The GZ mergers catalogue contains 3003 systems identified as mergers.  Of these, 1070 are potential candidates for our study, where the system lies within the main SDSS footprint (blue area in Figure~\ref{cover}), and at least one of the objects in the system is identified in 2MASS at a magnitude of $K<15.36$ (if both objects in a system are identified in 2MASS we require that both have $K<15.36)$.  

Of the 1070 GZ identified systems, 104 contain two individual galaxies in our 2MASS sample.  However, only 76 of these 104 meet our initial selection criteria ($\Delta V < 500~$km~s$^{-1},~5~h^{-1}<r_{\rm{sep}}~< 20~h^{-1}$~kpc, $K<15.36$).   Of the remaining 986 GZ systems, 825 have single object counterparts in 2MASS (the other 141 systems are fainter than our magnitude cut).  

We use the stellar mass estimates of \citet{Darg10a} to determine which of these systems could be classified as major mergers.  Of the 825 single 2MASS objects, 283 are double-redshift systems that meet our initial criteria, but just 26 meet all our selection criteria for major mergers.  An additional 497 of systems appearing as single objects in 2MASS are single-redshift systems that meet our initial selection criteria (except velocity difference).  However, just 23 of these systems meet all the criteria for major mergers.  

We add the 26 double-redshift major-merger systems to our main sample, and count the 23 single-redshift systems in the same way as described in Section~\ref{singlepair} for single-redshift pairs.  To extend this correction to the all-sky 2MASS sample, we identify the subset of GZ systems (14 total) that meet our criteria for major mergers with the brighter magnitude cut, then multiply their contribution to the pair fraction by the area of the all-sky sample divided by the area of the SDSS sample.  

Based on this analysis, we find that roughly half of the systems that would have been included in our initial selection are blended into single objects in the 2MASS catalogues.  The effect is less prominent in terms of total number of major-merger pair candidates.  The comparison with the GZ catalogues adds a total of 49 candidate pairs (26 double-redshift, 23 single-redshift) to our initial selection of 178 major-merger pairs (137 double-redshift, 41 single-redshift).  

The 2MASS $K<15.36$ pair sample, after the inclusion of these blended pairs, includes 227 pairs of galaxies.  The increase of $\sim 33\%$ over the total number of pairs (170) in the KPAIR sample may be attributed to an increase in sky coverage of $\sim 20\%$, an overall increase in spectroscopic completeness of $\sim 10\%$, and perhaps a bit due to the use of the GZ mergers catalog to identify blended pairs, rather than a by eye determination.  In any case, stopping the blended pair analysis at this point we arrive at a pair fraction vs. $L_K$ that is only slightly higher than that presented in \citet{Xu12}.  

However, in the comparison described above, we also found a relatively large number of 2MASS selected pairs in our sample that are not identified in the GZ catalogue.  The initial 2MASS + SDSS sample contains 415 double-redshift pairs and 234 single-redshift pairs (initial cut only, without excluding minor mergers or primaries within one mag of the flux limit).  All of these objects appear in the SDSS catalogues, but the majority appear morphologically undisturbed.  Only 81 of these systems (56 double-redshift, 25 single-redshift) are picked up by the GZ morphology classification.   

To explore this issue further, we looked for pairs identified in the UKIDSS sample that were classified as single objects in 2MASS and do not appear in the GZ catalogues.  In Figure~\ref{blending}, we show one such system.  In total, we found 156 systems in the UKIDSS sample that (based on UKIDSS photometry) would meet the initial selection criteria for the 2MASS sample, but which are catalogued as single objects in 2MASS and are not in the GZ catalogue.  These pairs then constitute another population of merger candidates that must be accounted for in the 2MASS blending correction.

In total, we found 21 major-merger candidate pairs meeting all selection criteria ($\Delta V < 500~$km~s$^{-1},~5~h^{-1}<r_{\rm{sep}}~< 20~h^{-1}$~kpc, $K<15.36$) for the 2MASS sample (16 double-redshift, 5 single-redshift).  As with the GZ candidates, we include the double-redshift pairs in the pair fraction directly and treat the single-redshift pairs as described in Section~\ref{singlepair}.  We also account for the fact that UKIDSS only overlaps about $\sim 30\%$ of the 2MASS/SDSS area by multiplying the contribution of these new pairs by a factor of $1/0.3$.  

These corrections to account for blending in 2MASS comprise a substantial addition to the pair fraction, with the largest contribution coming at the high-mass end, where galaxies tend to be at higher redshifts (more likely to be close in projection and blended).  In terms of total numbers, the percent contribution to the pair fraction is $53\%$ double-redshift pairs, $4\%$ single-redshift pairs, $16\%$ pairs added from GZ, and $27\%$ from pairs in UKIDSS that are missed in both 2MASS and GZ.   

Thus, in the 2MASS $K<15.36$ sample, the correction for blended pairs constitutes $43\%$ of the measured pair fraction, up from $\sim 15\%$ in the KPAIR sample, which accounts for most of the increase in the pair fraction we measure compared to that presented in \citet{Xu12}.  When blended pairs are accounted for in the 2MASS sample in this manner, the resulting pair fraction is in good agreement with that determined from the UKIDSS sample.  This analysis highlights the fact that 2MASS photometry is only useful as a photometric catalog for pair selection when combined with higher quality photometry (SDSS and UKIDSS in this case) to assess the effects of blending.  While we believe the UKIDSS sample is the most robust for calculating the low-redshift pair fraction, we also include the 2MASS samples in all the following analyses for comparison.

\subsection{Pair Velocity Distribution}

\begin{figure}
\begin{center}
\includegraphics[width=90mm]{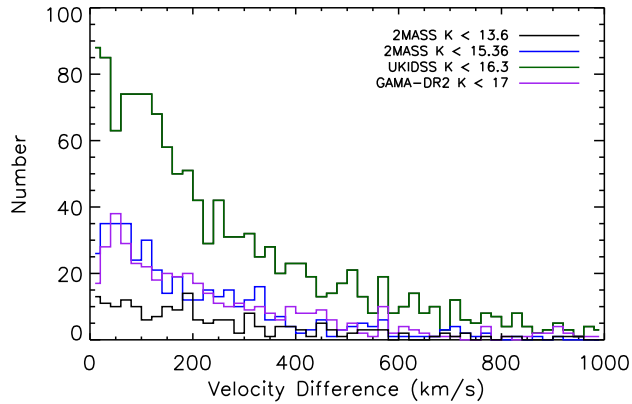}
\caption{\label{dv} Velocity difference of pairs in the spectroscopic sample.}
\end{center}
\end{figure}

In Figure~\ref{dv}, we show the  distribution in velocity difference for pairs in the various samples.  While the majority of pairs lie at velocities differences below $500$~km~s$^{-1}$, roughly $15\%$ are at higher velocities, up to an initial selection criterion of $< 1000$~km~s$^{-1}$.   In the analysis that follows, we only consider pairs at radial velocity differences of $< 500$~km~s$^{-1}$.  

\section{Pair Fraction and Merger Rate}
\label{pairfrac}

\begin{figure}
\begin{center}
\includegraphics[width=90mm]{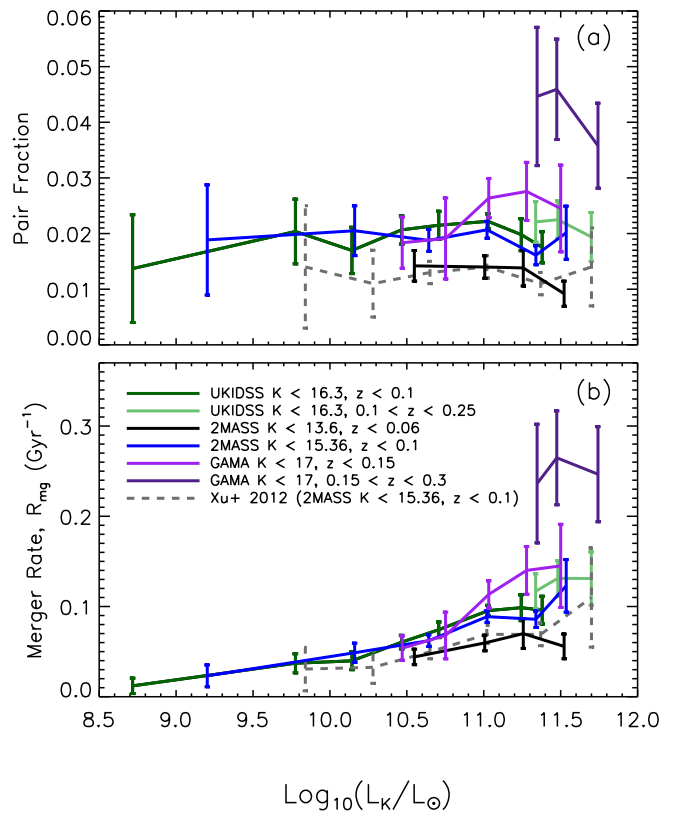}
\caption{\label{pf}  (a) The major-merger pair fraction as a function of $K-$band luminosity for samples in this study.  The error bars show $1~\sigma$ Poisson counting errors (based on small number statistics from \citealt{Gehr86} where appropriate).    The dashed line shows the low-redshift pair fraction  derived by \citet{Xu12} in their KPAIR sample, combining 2MASS photometry with SDSS-DR5 spectroscopy.  We split our samples into various redshift bins, which are denoted in the plot.  In general, we find a higher pair fraction than that derived by \citet{Xu12}.  This is due mostly to the more thorough treatment of blended objects in 2MASS, but also to higher spectroscopic completeness.  The UKIDSS sample covers the largest volume and should be the best estimate of the true low-redshift pair fraction.  (b) The merger rates (per Gyr per galaxy) implied by assuming the stellar-mass-dependent timescale and conversion from pair fraction given in equations~\ref{tmgeq}~and~\ref{rmgeq}, and that 100\% of pairs eventually merge.}
\end{center}
\end{figure} 

\begin{figure}
\begin{center}
\includegraphics[width=90mm]{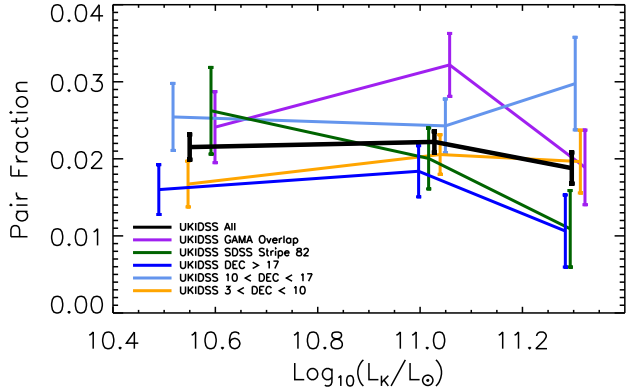}
\caption{\label{cvpf}  The field to field variation in measured pair fraction over five subregions of the UKIDSS sample.  In each case the pair fraction is measured in the same three luminosity bins for all galaxies at $z<0.25$.  The full UKIDSS sample is shown in black, the equatorial strip including all three GAMA fields in purple ($-2 < \rm{DEC} < 3~\rm{and}~129 < \rm{RA} < 223.5$), the SDSS Stripe 82 strip in green ($-2 < \rm{DEC} < 2~\rm{and}~310 < \rm{RA} < 60$, opposite the GAMA fields on the sky), and three other strips at the various declinations listed in the plot.  We find a $\sim 25\%$ dispersion in the central luminosity bin ($10.8 < \rm{Log_{10}}$$(L_K) < 11.2$, for which we will later measure the evolution with redshift).  The dispersion may be attributed to $\sim 15\%$ statistical error and $\sim 10\%$ systematic due to cosmic variance.}
\end{center}
\end{figure}

Here we consider only major-merger candidate pairs ($L_{\rm{primary}}/L_{\rm{secondary}} < 10^{0.4}$)  for which the primary (more luminous) galaxy is at least one magnitude brighter than the selection limit of the sample to ensure completeness in detecting major-merger secondaries.  We also imposed a projected separation criterion of $5 < r_{\rm{sep}} < 20~h^{-1}$ kpc, and a velocity difference upper limit of $500$~km~s$^{-1}$ for the reasons described in Section~\ref{pairs}.   

To compute the pair fraction, we add the total number of galaxies in double-redshift pairs, where all selection criteria are met, to the sum of the single-redshift pair probabilities, where single-redshift pair candidates meet all selection criteria except velocity difference.   Thus, the pair fraction is

\begin{equation}
f_{\rm{pair}}=\frac{N_{\rm{2z}} + \sum\limits_{i=1}^n P_{\rm{1z}}(i)}{N_{\rm{tot}}},
\end{equation}

where $N_{\rm{2z}}$ is the number of double-redshift pair galaxies meeting all the selection criteria,  $n$ is the number of single-redshift pairs, $P_{\rm{1z}}(i)$ are the probabilities associated with each of the single-redshift pair galaxies, as derived in Section~\ref{singlepair}, and $N_{\rm{tot}}$ is the number of galaxies in the full sample(with spectroscopic redshifts) that are $>1$ mag brighter than the selection limit, and fall within the redshift/luminosity range being considered.   For the 2MASS samples, the pair fraction includes the pairs identified via comparison with GZ and UKIDSS, as described in Section~\ref{blendcorr}.  

In this calculation of the pair fraction, only galaxies with spectroscopic redshifts are counted in ``$N_{\rm{tot}}$".  Thus, we are implicitly assuming that, after the single-redshift pair correction is made, the spectroscopic completeness for pair galaxies is equal to that of the full sample.  In other words, the drop in completeness at close angular separations seen in Figure~\ref{pairbias} is accounted for by the single-redshift pair correction, such that the spectroscopic completeness for pairs is flat at the square of the general level of of completeness for each sample as a function of angular separation.

In Figure~\ref{pf}a, we show the pair fraction in a variety of redshift and luminosity bins for the four samples considered here (error bars show $1~\sigma$ Poisson counting errors).  The corresponding merger rates are shown in Figure~\ref{pf}b and are discussed further in Section~\ref{mergerrate}.  We compare these results with those of \citet{Xu12} for their ``KPAIR" low-redshift sample, which was selected using 2MASS + SDSS DR5 with the same criteria.  

In the 2MASS $K<15.36$ sample, we find a higher pair fraction of $\sim 2 \%$ compared to the $\sim 1.4 \%$ found by \citet{Xu12} in their consideration of essentially the same sample.  This difference is mostly due to the more thorough treatment of the issue of blending in 2MASS described in
Section~\ref{blendcorr}, but also likely due, in part, to the higher spectroscopic completeness in our sample.   Accounting for blended pairs in the 2MASS sample brings the $K<15.36$ pair fraction into good agreement with that derived from the UKIDSS sample.  

\subsection{The Dwarf Galaxy Pair Fraction}

 With the low-redshift UKIDSS sample, we push more than an order of magnitude lower in stellar mass than previous studies.  We find no evidence of a higher pair fraction among low-mass galaxies.  However, we are only considering major mergers here, while the local group study of \citet{Fatt13} that found a $30\%$ pair fraction was considering all pairs separated by up to 3 magnitudes in brightness.  
 
The local group sample of \citet{Fatt13} contains many galaxies that would be beyond our detection limits, so we are not able to make a direct comparison with their study.  However, galaxies pairs in the mass range of the Magellanic clouds would be detectable in the UKIDSS sample out to $\sim 50$~Mpc.  While we cannot probe a dynamic range of 3 magnitudes brightness difference and still ensure complete detection of secondaries, we can put a lower limit on the pair fraction by considering the total fraction of galaxies in any pairs separated by up to 3 magnitudes and with a primary mass less than $10^{9.5}~M_\odot$.  

In the UKIDSS sample (which contains $\sim 1000$ dwarf galaxies), we find only 4 double-redshift pairs, and 32 single-redshift pairs.  Given that these single-redshift pairs are all at very low redshifts, they are highly unlikely to be physical pairs (see Figure~\ref{probmap}).  Even if all these galaxies were in physical pairs, it would only amount to a pair fraction of $\sim 6\%$.  The 2MASS samples contain fewer low-mass galaxies, but also feature a relatively low dwarf pair fraction.  

Thus, we can conclude that the pair fraction among dwarf galaxies is at least $1-2\%$, but we do not find evidence here to suggest the $\sim 30\%$ dwarf pair fraction found by \citet{Fatt13} in the local group is a general trend in the local universe.  

\subsection{Evolution Versus Sample Variance}
\label{cosvar}
We expect that the UKIDSS $z<0.1$ sample is the best current estimate of the pair fraction as a function of $K-$band luminosity in the local universe, given the depth and diversity of sight lines on the sky of this sample.  The UKIDSS sample appears to indicate that the pair fraction remains flat at $\sim 2\%$ over the entire luminosity range considered and out to redshift $z=0.25$. Interestingly, however, the 2MASS $K<13.6$ all-sky pair fraction remains significantly lower even after correction for blended pairs, while the GAMA sample shows a slightly higher pair fraction for bright galaxies in the $z<0.15$ bin, and a much higher pair fraction in the $0.15<z<0.3$ bin.

These results would appear to indicate that the pair fraction of massive galaxies may be increasing by a factor of $3$ or more from $z\sim0$ to $z\sim0.3$ (the redshift range bracketed by the 2MASS and GAMA samples).  This would imply a rather extreme evolution of the merger rate of galaxies, with $m\sim 4$ in the relation $f_{\rm{pair}} \propto (1+z)^m$.  
However, these results could also simply be due in part to small number counting statistics and in part to so-called ``cosmic variance" due to large-scale structures in the volumes surveyed, as the actual fraction of physical and projected pairs has been shown to depend on large-scale environment \citep{Lin10, Dera11, Jian12}

To investigate these effects, we recalculated the pair fraction of UKIDSS galaxies in five subsamples for all galaxies at $z<0.25$.  The results of this exercise are shown in Figure~\ref{cvpf}, where the result for the full UKIDSS sample is shown in black, for an equatorial strip containing the GAMA fields in purple ($-2 < \rm{DEC} < 3$ and $129 < \rm{RA} < 223.5$), for the SDSS Stripe 82 equatorial strip in green ($-2 < \rm{DEC} < 2~\rm{and}~310 < \rm{RA} < 60$, opposite the GAMA fields on the sky), and for three other strips at the various declinations listed in the plot.  Focusing on the luminosity bin $10.8 < L_{\rm{K}} < 11.2$ (which we will later use to study evolution of this quantity), these results indicate a $\sim 25\%$ dispersion in the pair fraction measured in different directions on the sky in the UKIDSS sample.  Poisson (counting) error should be at the $\sim 14\%$ level given the average number of pair galaxies per subsample in this bin is $\sim 50$.  If Poisson error and cosmic variance are the dominant contributions to the $25\%$ mentioned above they should add in quadrature, implying cosmic variance at the $\sim 20\%$ level in each subsample.

\citet{Driv10} showed that cosmic variance, $\sigma_{\rm{cv}} \propto 1/\sqrt{N}$, where $N$ is the number of independent sight lines.  In the full UKIDSS sample, we expect both the counting error and cosmic variance to be reduced by a factor of $\sqrt{5}$ from the jack-knife resampling values mentioned above, given that the number of objects per bin goes up by a factor of $5$, and with $5$ sight lines, the cosmic variance should be reduced by a factor of $\sqrt{5}$.  Thus, the counting error for the full sample is at the $\sim 6\%$ level, and cosmic variance at $\sim 9\%$, for a total error of $\sim 11\%$ once added in quadrature.

\citet{Lope14} studied cosmic variance in pair studies, using data from the ALHAMBRA\footnote{http://alhambrasurvey.com/?lang=en} survey, as a function of the number density of the parent samples, and the volume surveyed.  They found that the uncertainty due to cosmic variance may be expressed as 

\begin{equation}
\label{cveq}
\begin{split}
\sigma_{\rm{cv}}(n_1,n_2,V) = 0.48 \times \left( \frac{n_1}{10^{-3}~\rm{Mpc}^{-3}}\right)^{-0.54}\\
\times \left(\frac{V}{10^5~\rm{Mpc}^3} \right)^{-0.48}\times \left(\frac{n_2}{n_1}\right)^{-0.37},
\end{split}
\end{equation}

where $n_1$ is the number density of the parent sample for primary galaxies, $n_2$ is the number density of the parent sample for secondaries (i.e., 1 mag fainter than for primaries), and $V$ is the volume surveyed.    If we compute $\sigma_{\rm{cv}}$ for the full UKIDSS sample directly from equation~\ref{cveq}, we arrive at a result of $\sigma_{\rm{cv}} \approx 4\%$, roughly half our previous estimate.  The ALHAMBRA survey consists of many small pencil beam sight lines, sampling a quite different volume than that of our UKIDSS sample, and, thus, equation~\ref{cveq} may not be appropriate in this case, but we make mention of this apparent agreement here given that the direct comparison may be made.

We note the highest pair fraction in UKIDSS is measured in the direction of the GAMA fields, which also happen to overlap with the so called ``Sloan Great Wall" \citep{Gott05}.    In addition, the all-sky 2MASS $K<13.6$ pair fraction remains low, even after the blending correction.  Several recent studies (e.g., \citealt{Keen10a, Keen12, Keen13, Whit14}) have found the local universe appears under-dense at $z\lesssim0.07$.  \citet{Dera11} have shown that the pair fraction increases with local galaxy space density,  and, thus, we suspect the higher/lower pair fractions measured in these volumes may be reflective of these observed  over/under-densities.  

Calculating estimates for the other samples in the same way as presented above, we find that the combination of counting error and cosmic variance should be $\sim 20\%$ for our GAMA sample, $\sim 22\%$ for the 2MASS $K< 13.6$ sample, and $\sim 15\%$ for the 2MASS $K < 15.36$ sample.  Given these analyses, we conclude that the scatter in Figure~\ref{pf}a is consistent with the expected dispersion due to counting errors plus cosmic variance due to large-scale structures in the various survey volumes.  The full UKIDSS sample, where we find a relatively flat pair fraction at $\sim 2\%$ as a function of luminosity, covers the largest volume and should be the most robust to cosmic variance ($\sigma_{\rm{cv}}\sim 9\%$).

\subsection{The RCS1 sample}

To measure evolution in the pair fraction we compare with a stellar-mass-selected pair sample from the RCS1 \citep{Glad05}.   The RCS1 samples roughly $33$~deg$^2$ on the sky, and is comprised of 10 widely separated regions.

The observations, data reduction, and photometric redshift determinations for the RCS1 sample are presented in \citet{Hsie05}.   Luminosity-selected galaxy pairs from the RCS1 sample were presented in \citet{Hsie08}.  For the present study, B.~C.~Hsieh performed a reanalysis of the RCS1 close pair sample based on a stellar mass selection and other criteria designed to match our low-redshift selection from 2MASS and UKIDSS.     The details of the pair selection methods, completeness and projection corrections, etc., are nearly identical to those described in \citet{Hsie08}, except that the fundamental selection is by stellar mass, rather than rest-frame $R-$band luminosity.  Here we briefly describe these methods, but for a more comprehensive description we refer the reader to \citet{Hsie08}.  

\citet{Hsie08} and \cite{Glad05} describe the calibration of RCS1 using SDSS stellar photometry as a standard catalog.   Calibrations were performed to ensure consistent photometry from field to field, from pointing to pointing within a given field, and from chip to chip on the CCD.   Here we are measuring the pair fractions in galaxies that are detected at very high signal to noise (minimum $S/N \sim 10$, but much higher for the vast majority of sources) in the RCS1 sample, and, thus, we do not expect photometric anomalies to be a significant source of uncertainty in this sample.  

The RCS1 photometric redshifts are based on four band photometry in the  $B, V, R_c$, and $z^{\prime}$ bands.  While one might worry about the quality of four-band photo-z's used for the purpose of selecting physical pairs of galaxies, it turns out to be a non-issue, given that objects of similar brightness that are close in angular separation on the sky have a very high probability of being physically associated (e.g., as shown in Figure~\ref{probmap}).  \citet{Hsie08} demonstrate that even after discarding the redshift information altogether it is possible to perform a robust study of pair galaxies using the appropriate completeness and projection corrections (see Figure 10 from \citealt{Hsie08}).   Following the analysis of \citet{Hsie08}, we search for companion galaxies over the redshift range $z_{primary} \pm n\sigma_z$, where $n=2.5$ and $\sigma_z$ is the $68\%$ redshift uncertainty interval.   

We calculate a projection correction for each RCS1 galaxy individually by calculating the mean surface density of all objects near a candidate pair galaxy that satisfy the criteria $\Delta z \leq n\sigma_z$ and $\Delta \rm{Log_{10}}(M^*) \leq 0.4$, multiplied by the search area for companions $(5-20~h^{-1}$ kpc).  This value is then subtracted from the actual number of companions found for any candidate primary galaxy.

We also adopt reliability corrections taken from \citet{Xu12} to account for clustering and pairs with $\Delta V > 500$~km/s (predominantly in cluster environments).  \citet{Xu12} found that a flat correction may be applied across all redshift bins of $0.94$ to correct for clustering, and $0.91$ for pairs with $\Delta V > 500$~km/s.  These are both multiplicative corrections, and, thus, serve to reduce the estimated pair fraction in each redshift bin by a factor of $0.86$.

In addition, we restrict the parent sample to include only galaxies meeting a redshift quality criterion of $\sigma_z / (1+z) \leq 0.3$.  This criterion was determined by \citet{Hsie08} as optimal  for including the maximum fraction of the data, while simultaneously minimizing the noise in the measurement due to poor photo-z's.  We then determine a completeness correction factor for each candidate primary galaxy by determining the ratio of the total number of galaxies in a 0.1 mag bin centered at the $R_c$ magnitude of that galaxy divided by the number of galaxies in that bin satisfying the redshift criterion (typical completeness correction factors are $\sim 1.1-1.2$).  

Another source of incompleteness in the RCS1 sample is missing blended pairs.  The median seeing for RCS1 observations was $\sim 0\farcs9$, corresponding to a physical size of $2.5~h^{-1}$~kpc at the low-redshift limit of this study ($z=0.25$), and a size of $4.8~h^{-1}$~kpc at the high-redshift limit ($z=0.8$).  Thus, with our minimum separation criterion of $r_{\rm{sep}} > 5~h^{-1}$~kpc, it is clear that some pairs will be missed due to blending, particularly on the high-redshift end of the survey.  \citet{Hsie08} estimated what they called a ``seeing correction" to account for this effect by comparing pair fraction estimates from the best and worst seeing conditions observed.   They found that it was not a significant source of bias.  \citet{Xu12} estimated the blending correction directly using high resolution $HST-$ACS imaging of the COSMOS field, and found the fraction of pairs missed due to blending was $[0.01, 0.06, 0.08, 0.2]$ in their four redshift bins over the range $0.2 < z < 1$.  We interpolate these values to our own redshift bin centers to infer corrections of $[0.03, 0.07, 0.08, 0.11]$ for the four RCS1 bins over the range $0.25 < z < 0.8$.

Thus, in addition to our own calculated corrections for incompleteness (due to photo-z error) and chance projection, we adopt the corrections made by \citet{Xu12} for clustering, $\Delta V < 500$~km/s, and blended pairs.  The combination of the three corrections taken from \citet{Xu12} results in a reduction of the estimated pair fraction in the four RCS1 redshift bins by factors of $[0.88, 0.91, 0.93, 0.97]$, respectively, from low to high redshift.

We computed stellar masses for the RCS1 sample using SED fits to optical data (\citealt{Bruz03} templates with a Salpeter IMF), which is not strictly equivalent to stellar masses estimated via $K-$band luminosity.  To test how these two quantities are related, we first computed stellar masses for all UKIDSS pair galaxies plus a selection of $\sim 5000$ random galaxies in the UKIDSS sample using four bands of SDSS photometry and the same methods used for the RCS1 sample.  This comparison yielded the result that stellar masses computed with SDSS photometry and spec-z's are $\sim 0.13$ dex higher (with $\sim 0.25$ dex scatter) than $K-$band luminosities for the same objects.   This result is in good agreement with other estimates of $M/L_K$ using a Salpeter IMF (e.g., \citealt{Cole01}).  

We also compared RCS1 stellar masses and photo-z's directly with UKIDSS $K-$band luminosities in fields where the RCS1 and UKIDSS-LAS overlap (for several hundred objects with available spec-z's).  We found that RCS1 photo-z's are in good agreement with spectroscopic redshifts (median $\Delta z / (1+z) \sim 0.03$).  We also found that RCS1 stellar masses at $z < 0.15$ are $\sim 0.13$ dex higher than $K-$band luminosities for the same objects (consistent with the previous comparison).  However, we noted a downward trend in the $M/L_K$ ratio toward increasing redshift with RCS1 estimated stellar masses being $\sim 0.2$ dex lower than $K-$band luminosities by $z \sim 0.5$.  This result is qualitatively consistent with the expectation of stellar mass increasing relative to rest-frame $K-$band luminosity toward $z=0$ as star formation decreases and galaxies evolve passively.

\subsection{Pair Fraction Evolution}
\label{mergerrate}
\input{Table1}
\input{Table2}

\begin{figure}
\begin{center}
\includegraphics[width=90mm]{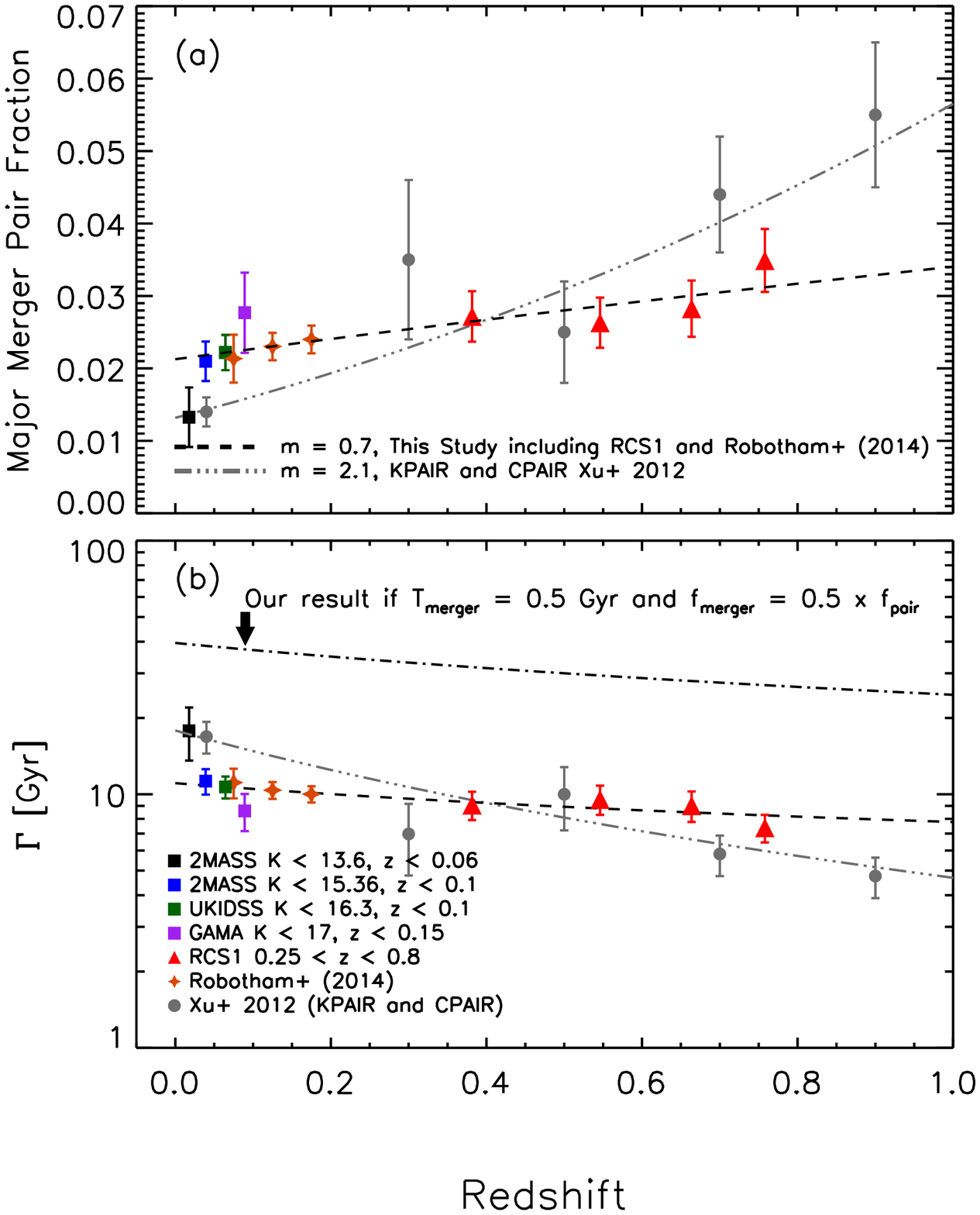}
\caption{\label{pfvz} (a) Pair fraction for major mergers with primary galaxies in the luminosity range $10.8 < \rm{Log_{10}}$$(L_K) < 11.2$ as a function of redshift.  Colored squares show results from this study.  Red triangles show the results using a stellar mass selection in the RCS1 sample that matches the number density in the UKIDSS sample.   Orange diamonds show the results from \citet{Robo14}.  Gray circles show the results of \citet{Xu12}.   Error bars in all cases include statistical errors plus an estimate of the systematic due to cosmic variance.   Fits are shown assuming the pair fraction evolution may be parameterized as $f_{\rm{pair}} \propto (1+z)^m$.  The dash-triple-dot line shows the result of \citet{Xu12}, with $m=2.1$.  The dashed line shows the result of $m=0.7\pm0.1$ combining our low-redshift data with the RCS1 pair sample, and the results of \citet{Robo14}.  (b) The inverse of the merger rate ($\Gamma = 1/R_{\rm{mg}}$), showing the typical time between major merger events for a galaxy, as a function of redshift.    The fits are the same shown in panel (a), now converted to $\Gamma$.  The dashed line shows the result from this study using the conversion from pair fraction to merger rate shown in Equations~\ref{tmgeq}~and~\ref{rmgeq} (assuming all pairs merge).  The dash-dot line shows our result if we assume a merger timescale of $0.5$ Gyr, and that only 50\% of pairs merge.}
\end{center}
\end{figure} 

\begin{figure}
\begin{center}
\includegraphics[width=90mm]{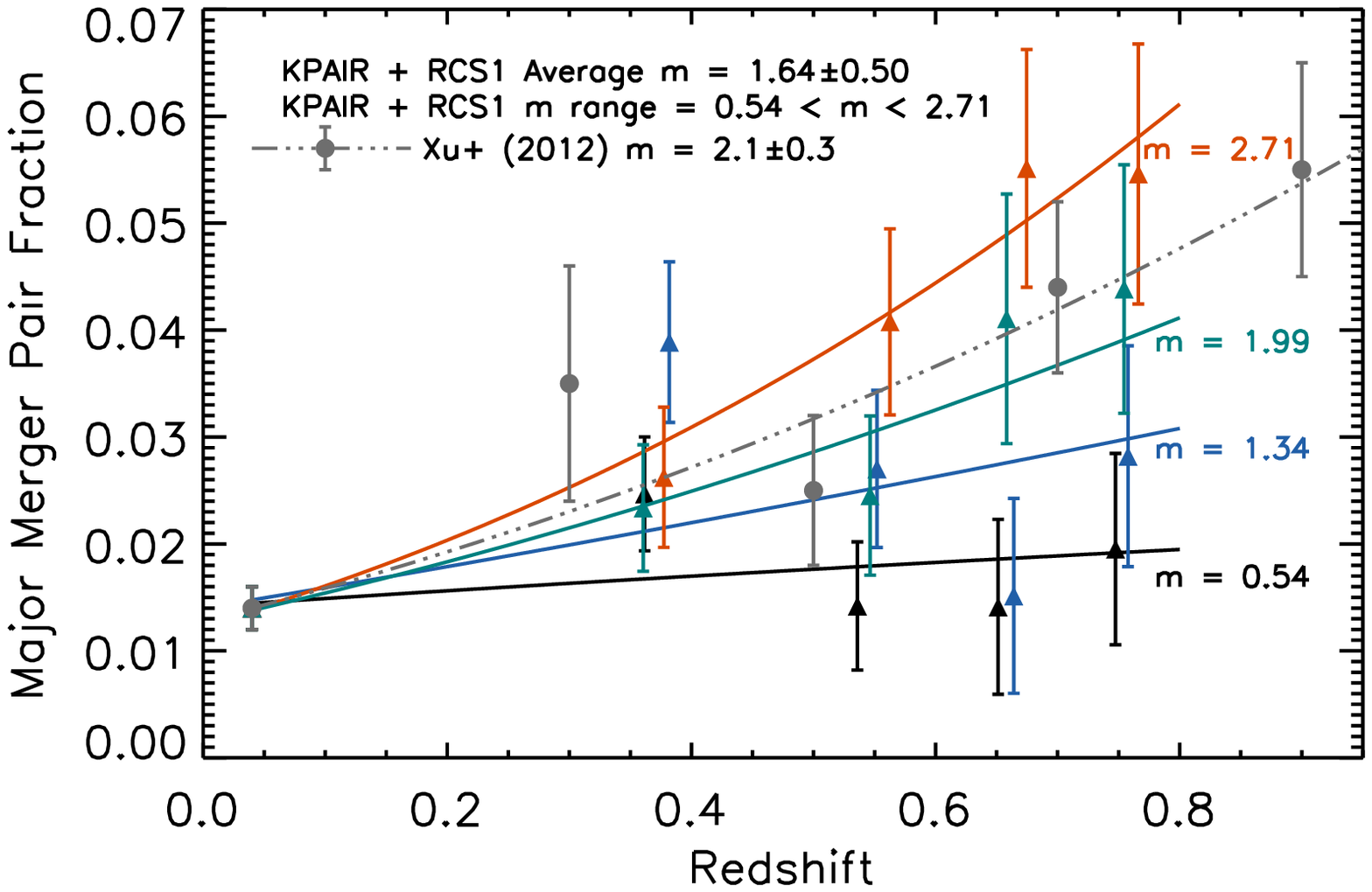}
\caption{\label{rcs1} Sample variance estimate for the pair fraction evolution using the RCS1 data.  Each set of colored triangles and corresponding solid line represents a pair fraction evolution measurement in four redshift bins of equal volume covering $0.25<z<0.8$ over one COSMOS-sized subfield of $1.7$~deg$^2$ in the RCS1.  In each RCS1 bin we choose a stellar mass range (0.4 dex wide) that matches the number density of the UKIDSS sample at low redshift.  We combine these measurements with the low-redshift KPAIR result and fit the evolution assuming $f_{\rm{pair}} \propto (1+z)^m$ to investigate cosmic variance in the RCS1 results.   We made this measurement over 20 separate subfields, but here we only show four examples to display the range of $m$ values obtained (min, max, and two mid-range results are shown).    For KPAIR + RCS1 we find an average value for the exponent of $m=1.64\pm 0.5$.  The filled circles and dash-triple-dot line shows the $m=2.1\pm0.3$ result from KPAIR + CPAIR from \citet{Xu12}.  We note that the individual RCS1 fields show evolution ranging from steeper than CPAIR ($m=2.7$) to much flatter ($m=0.5$).}
\end{center}
\end{figure}

The evolution of the pair fraction is commonly parameterized as $f_{\rm{pair}} \propto (1+z)^m$.  In general, the most difficult issue in comparing results from different studies of the evolution of the galaxy pair fraction is that selections and methods are often quite different from one study to the next.  Our selection and methods are very similar to those of \citet{Xu12}, and so we first make a comparison with their results.   \citet{Xu12} used a stellar mass bin of $10.6 < \rm{Log_{10}}$$(M_K) < 11$ to track the pair fraction in $L\sim L^*$ galaxies at $z < 1$.  Through matching the number densities of objects in our 2MASS $K<15.36$ sample to those of  \citet{Xu12}, we find that this stellar mass bin is equivalent to a $K-$band luminosity selection of  $10.8 < \rm{Log_{10}}$$(L_K) < 11.2$.

In Figure~\ref{pfvz}a, we show the pair fraction as a function of redshift.  The dash-triple-dot line shows the pair fraction evolution ($m=2.1\pm0.3$) determined by \citet{Xu12}, from the combination of their low-redshift KPAIR spectroscopic sample from 2MASS + SDSS-DR5 with the higher-redshift CPAIR photometric sample.   This relatively steep evolutionary trend is driven largely by the low-redshift anchor point derived from their KPAIR sample.  

We have shown that, after more careful consideration of blended pairs, the best estimate for the major-merger pair fraction using 2MASS + SDSS is closer to $2\%$. This is then in agreement with the result from the UKIDSS sample, which is deeper and does not suffer from blending of pairs (given the criterion $r_{\rm{sep}} > 5~h^{-1}$~kpc).

The red triangles in Figure~\ref{pfvz} show the results from the RCS1 sample, where we have matched the UKIDSS number density ($\sim 7.5 \times 10^{-4}~$Mpc$^{-3}$) in four redshift bins of equal volume over the range $0.25 < z < 0.8$.   We find the pair fraction to be relatively flat at $\sim 3 - 3.5\%$ over this redshift range.  These measurements are in good agreement with the CPAIR sample, although combined with our low-redshift pair fraction measurement, yield a flatter evolution with $m\approx 0.7$.    

We also compare with the recent results from \citet{Robo14} based on the GAMA survey (orange stars in Figure~\ref{pfvz}).  They measured the pair fraction in three redshift bins over the range $0.05 < z < 0.2$.  The selection they used for this measurement was a mass ratio of $1:3$ and $r_{\rm{sep}} < 20~h^{-1}$ kpc.  We correct their values down by $5\%$ to match our projected separation range of $5 < r_{\rm{sep}} < 20~h^{-1}$ kpc \citep{Patt97, Patt00} and down by another $2.5\%$ to accommodate the change from a $1:3$ mass ratio to $1:10^{0.4}$ (calculated empirically using our UKIDSS sample).  

The fit shown as a dashed line in Figure~\ref{pfvz} includes our pair fraction measurements from 2MASS ($K<15.36$), UKIDSS, GAMA, and RCS1 as well as the results from \citet{Robo14}.   We measure an exponent of $m= 0.7\pm 0.1$, and we find that the results from \citet{Robo14} are in such excellent agreement with our own that adding or omitting their data from our evolution fit leaves the measured exponent essentially unchanged.  In their own study, \citet{Robo14} found $m=1.53 \pm0.08$, but they were pulled toward this higher value by previous low-redshift estimates, including KPAIR, as well as some the of higher pair fraction estimates from various studies in the COSMOS and other fields.   Beyond $z=0.2$, there is still significant uncertainty in the pair fraction, but the RCS1 sample improves this measurement significantly.  

\subsection{Cosmic Variance in the Pair Fraction Evolution Results}

To explore the issue of cosmic variance and the comparison between the CPAIR and RCS1 samples, we performed a jack-knife resampling of the RCS1 data by making the same pair fraction evolution measurement on 20 COSMOS-sized fields of $1.7$ deg$^2$ each.  We then fit each result including the low-redshift KPAIR measurement and assuming $f_{\rm{pair}} \propto (1+z)^m$.  We performed this resampling for both the fixed stellar mass bin matched to the CPAIR density ($10.5 < \rm{Log_{10}}$$(M_K) < 10.9$) and for the variable mass binning scheme designed to match to the UKIDSS density (see Table 2).  Both methods yielded very similar results.  

In Figure~\ref{rcs1}, we show a selection from the resampling results for four of the RCS1 subsamples (colored triangles with solid lines) alongside the result from \citet{Xu12}.  These four examples show the minimum and maximum $m$ values measured, as well as two intermediate values.  The RCS1 sample is not as deep as the CPAIR sample, so we can only directly make this comparison out to $z=0.8$.  Using the combination of KPAIR plus the 20 RCS1 subsamples, we find a mean value for the exponent of $m=1.64\pm0.1$.  However, we note that some of the RCS1 subsamples yield an even higher $m$ value than CPAIR (highest value $m\sim2.71\pm0.4$) and some a much lower value (lowest $m\sim 0.54\pm 0.4$).  

The rms error implied by the jack-knife resampling of the RCS1 suggests $\sim30\%$ per bin in each subsample (cosmic variance combined with Poisson error).   Much of this scatter is most likely coming from small number statistics ($\sim20\%$ inferred counting error per subsample bin).  Photometry and completeness issues could be affecting the results in the highest RCS1 bin (where we find a somewhat larger scatter of $\sim 40\%$), but this should not be present in the three lower redshift bins because the photometry is uniform, the targets at very high signal to noise, and completeness high.  Thus, we infer a systematic due to cosmic variance of $\sim 20\%$ per subsample bin in this resampling of RCS1.  

Applying equation~\ref{cveq} to the full RCS1 sample, we find an estimated $\sigma_{\rm{cv}}\approx 3-4\%$.  The counting error per bin in the full RCS1 sample is $\sim 6\%$, and estimating the cosmic variance in the full sample given that the resampling was performed over the 10 independent RCS1 sight lines we find $\sigma_{\rm{cv}} \approx 9\%$, again roughly a factor of two or three higher than the estimate from equation~\ref{cveq}.

\subsection{Merger Rate Evolution}

The differential galaxy merger rate ($R_{\rm{mg}}$) is the probability for a galaxy to be involved in a major merger per Gyr.  Thus, $R_{\rm{mg}} \propto f_{\rm{pair}}/T_{\rm{mg}}$, where the merger timescale, $T_{\rm{mg}}$, must be assumed to calculate the merger rate.  Here, we first assume the same merger timescale as \citet{Xu12}, which they arrived at by combining the simulation results of \citet{Kitz08} with those of \citet{Lotz10}.  The final form of the merger timescale is: 

\begin{equation}
\label{tmgeq}
T_{\rm{mg}}=0.3{\rm{Gyr}} \times \left( \frac{M_{\rm{star}}}{10^{10.7}M_{\odot}} \right)^{-0.3} \left(1+\frac{z}{8}\right).
\end{equation}

To use this equation we must assume a conversion factor to go from $K-$band luminosity to stellar mass.  \citet{Xu12} assume $M_{\rm{stars}}/L_{K} = 0.54$ (for a Chabrier initial mass function).  A detailed comparison of their results with our 2MASS $K<15.36$ sample indicates a conversion factor of $M_{\rm{stars}}/L_{K} \approx 0.63$ (0.2 dex), which we use to convert our measured $K-$band luminosities to stellar masses equivalent to those used by \citet{Xu12}.  This equation yields a merger timescale of $T_{\rm{mg}}\approx 0.3~$Gyr for galaxies at $M\sim M^*$.  The differential merger rate is then 

\begin{equation}
\label{rmgeq}
R_{\rm{mg}}=A \times f_{\rm{pair}}/T_{\rm{mg}}, 
\end{equation}

where $A=1.19$ to convert from a minimum mass ratio of $1/3$ used in the simulations, to that of $1/10^{0.4}$ used here (see \citealt{Xu12} for details).  Here we are assuming all pairs eventually merge.

In Figure~\ref{pf}b, we show the merger rate (assuming the above conversions from pair fraction to merger rate) as a function of $K-$band luminosity for the various samples considered in the pair fraction studies.   In Figure~\ref{pfvz}b, we show the inverse of the differential merger rate ($\Gamma = 1/R_{\rm{mg}}$) evolution as a function of redshift for the samples considered in this study as well as those from \citet{Xu12} and \citet{Robo14}.  This then shows how the typical time between merger events evolves for a galaxy.   

Again the dash-triple-dot line shows the evolution derived by \citet{Xu12}, and the dashed line shows the evolution inferred from our low-redshift sample combined with RCS1 and the results of \citet{Robo14}.  Similar to the pair fraction results, we find a less dramatic evolution in the merger rate than that implied by KPAIR + CPAIR.  However, in terms of integrated mergers per galaxy since $z=1$, we find a very similar result to that of \citet{Xu12}, of $\sim 0.8$ major mergers per galaxy since $z=1$.  

However, the merger timescale for $L\sim L^*$ galaxies derived using equation~\ref{tmgeq}~ is $\sim 0.3~$Gyr, which, according to recent work comparing simulations and local galaxy mergers, is probably close to the lower limit on the major merger timescale for galaxies meeting our selection criteria~\citep{Priv13}.  Furthermore, the assumption that $100\%$ of pairs merge is most certainly an upper limit.  Thus, the dashed and dash-triple-dot lines in Figure~\ref{pfvz}b effectively represent lower limits on the actual time between major mergers.  

For comparison, we also include in Figure~\ref{pfvz}b, an estimate of the time between major mergers  assuming the merger timescale is slightly longer ($0.5$~Gyr up from $\sim 0.3$~Gyr) and that only $50\%$ of pairs eventually merge.  The dash-dot line shows this result and emphasizes the strong dependence of the inferred merger history of galaxies on these parameters.  In the case of these more conservative assumptions, a typical galaxy would only undergo $\sim 0.2$ major mergers since $z=1$.  

Many studies from the literature of the merger rate evolution present a figure comparing the pair fraction or merger rate evolution as determined by studies using a variety of methods, or apply some conversion factors in order to combine results from different studies obtained using different methods to estimate evolution.  Here, we only make a direct comparison between studies that have been done using very similar selection criteria.  In the following section, we present a discussion of comparable studies in the context of this work.

\subsection{Discussion}
\label{discussion}
In the previous sections, we have shown that even studies using very similar selection criteria can arrive at rather different results when it comes to an estimation of the pair fraction of major mergers at any given redshift.  A meaningful comparison between studies becomes more difficult when the results are derived using different selections and methods, but here we discuss some recent studies which are related or comparable to this work.  

\citet{Lope12} studied stellar-mass-selected major-merger pairs ($M^* > 10^{11}, 10<r_{\rm{sep}}<30~h^{-1}$~kpc, mass ratio limit $1/4$) in the COSMOS field.  They further separate their sample into early and late-type galaxies (ETG/LTG).  They find that the evolution of the LTG pair fraction is much steeper ($m=4$) than that of ETGs ($m=1.8$).  This is in relative agreement with the trends found in other studies that separate ETG/LTG, or red/blue galaxies \citep{Lin08, Dera09, Bund09, Chou11, Lope11}.  Extrapolating to higher redshift, this result would appear to be in rough agreement with the relatively high pair fraction ($\sim 20\%$) among gas-rich galaxies (i.e., LTGs) found by \citet{Lope13}, although, again, these studies feature rather different selection criteria.

However, \citet{Lope12} also find that the fraction of ETG mergers is roughly twice that of LTG mergers, and their measured trends seem to indicate nearly all major mergers between massive galaxies at $z=0$ should be between ETGs.  This would seem to be in contradiction to the results of \citet{Kart07}, who (also studying pairs in the COSMOS field) state that the vast majority of merging systems at all redshifts in their sample are star-forming disk galaxies (down to $z=0.2$), or \citet{Chou12}, who find ``red-red" mergers are nearly absent in their sample at $z\sim 0.5$.  In addition, \citet{Depr10} find that dry mergers do not contribute significantly to the buildup of the red sequence at $z<0.7$.   While we do not attempt to quantify ETG/LTG fractions here, we can say that LTG major mergers are still a substantial fraction of the population of stellar-mass-selected major mergers at $z\sim0.1$ (both via visual inspection and comparison with the GZ catalogues).  

\citet{Dera09} find evidence that the pair fraction evolution depends on stellar mass (or luminosity) of the selected pairs, with a steeper evolution for lower stellar mass (or luminosity) galaxies.   This is consistent with the results presented by \citet{Hsie08} using the RCS1 sample.  \citet{Dera11} find that stellar mass, rather than luminosity, is the better indicator of a galaxy's merger history.   The highest stellar mass bin from \citet{Dera09} (Log$_{10}(M*)>10.5$) is closest to that of our own study ($10.6<$~Log$_{10}(M*)<11$) and they find an evolution with $m=0.51 \pm 2.01$, consistent (because of the large error bars) with essentially all results presented in this study.  

\citet{Depr07} determine a similar pair fraction at $z\sim 0.1$ to that found in our study ($\sim 2\%$), albeit with quite different selection criteria ($-21 < M_B< -18,~r_{\rm{sep}} < 20~h^{-1}~$kpc, and no mass ratio constraint).  Interestingly, they find essentially the same fraction by counting merger remnants identified by asymmetry, suggesting the timescale for post-merger asymmetry is similar to that of a $r_{\rm{sep}} < 20~h^{-1}~$kpc pair (and perhaps suggesting their asymmetry merger fraction is a quantity directly comparable to our own pair fraction measured here).   Combining their results with the higher redshift analysis of \citet{Lin04}, they find a relatively flat evolution of the merger rate with $m\approx1$, a value closer the result we get combining our own low-redshift results with RCS1.  

\citet{Depr05} use the same velocity and projected separation criteria as we have used here combined with a $B-$band luminosity selection.  They find the number of companions per galaxy $N_c = 0.0174\pm 0.0015$ and $N_c=0.0357\pm0.0027$ for the luminosity selections of $-22<M_B<-19, \langle z \rangle = 0.126$ and $-21<M_B<-18, \langle z \rangle = 0.126$, respectively.  These estimates could both be considered in agreement with our low-redshift samples given the considerations of cosmic variance detailed above.  

\citet{Patt08} find $N_c = 0.021 \pm 0.001$ in the SDSS for a sample selected by $5 < r_{\rm{sep}} < 20~h^{-1}~$kpc,$~\Delta V < 500$~km~s$^{-1},~-22 < M_r < -18$, and 1:2 luminosity ratio.  Applying the same criteria to a sample selected from the Millennium simulation \citep{Spri05}, the find $N_c = 0.0183\pm 0.0001$.  While the selection criteria are somewhat different, these values are both in relative agreement with our results for $L^*$ galaxies in our low-redshift sample.

\citet{Bund09} study stellar-mass-selected pairs at redshifts of $0.4 < z < 1.4$ over two fields totaling $\sim 320$ arcmin$^2$, and find that the pair fraction evolves with an exponent of $m=1.6\pm1.6$.  They conclude that major mergers alone cannot fully account for the  buildup of spheroidal galaxies since $z=1$.  Finding a steeper evolution (though still consistent with \citealt{Bund09}), and with the assumption of a particular galaxy stellar mass function, \citet{Xu12} contradict this result and conclude that major mergers are sufficient to account for the buildup of red quiescent galaxies and ellipticals.  They note that much of this discrepancy may be attributed to a factor of two difference in the assumed merger timescale in these two studies.   

\citet{Robo14} have recently published their pair fraction measurements from the GAMA survey.  Their data are drawn from three large equatorial fields totaling $\sim 144$ deg$^2$ with very high spectroscopic completeness to $R=19.8$.  These represent, by far, the most robust measurements of the pair fraction in the range $0.1 < z < 0.2$.  In a companion paper, \citet{Depr14} studied the characteristics of the GAMA merger sample in luminosity selected pairs to quantify the color, morphology, environment, and nuclear activity of merger candidates.

\citet{Robo14} combined their measurement of the pair fraction from the GAMA sample with other results from the literature at $z < 1$ and derived an exponent of $m=1.53\pm 0.08$.  However, it appears (in their Figure 15) that the measured evolution is made steeper by some of the low-redshift results from the literature (including KPAIR), and on the high-redshift end, the exponent gets pushed further up due, in large part, to the study from \citet{Kart07}.   Combining the results of \citet{Robo14} with our own results from 2MASS, UKIDSS, GAMA, and RCS1, we derive an exponent of $m=0.7\pm0.1$.  Our results are in such good agreement that, in fact, the exponent on the fit changes insignificantly whether or not we include the results from \citet{Robo14}.  

Thus, studies of major mergers at low-redshifts ($z < 0.2$) appear to be converging on a consistent result, while studies focused on constraining the evolution of the major merger rates of galaxies out to $z=1$ have yet to come to a consensus.  \citet{Lotz11} show that some of the discrepancies between studies can be accounted for once consistent assumptions for the merger timescale, as well as consistent sample selection and merger rate definitions are imposed.  

We believe the UKIDSS sample presented here represents the best current estimate of the low-redshift pair fraction, and that, when combined with the RCS1 sample, provides the best current estimate of the pair fraction evolution out to $z=0.8$.  With the assumption of different merger timescales and percentage of pairs that eventually merge, our results imply that $L^*$ galaxies have undergone $\sim 0.2-0.8$ major mergers since $z=1$.

\section{Summary}
\label{summary}

Here we have considered a large sample of close galaxy pairs effectively stellar-mass-selected in the $K-$band from UKIDSS and 2MASS photometry.   Combining these NIR photometry catalogues with available spectroscopic redshifts, we construct the largest complete spectroscopic sample to date of stellar mass selected galaxy pairs at relatively low redshifts.  Our pair sample is drawn from both wide and deep surveys, allowing us to investigate the close pair fraction and merger rates over a wider range in stellar mass  than was possible in previous spectroscopically complete studies.  

We find the pair fraction among major merger candidate galaxies to be flat as a function of stellar mass at $z\sim 0$, in contrast to results in the local group, which found a higher pair fraction among low mass galaxies.  We investigate the incidence of pairs in a wider range of mass ratio (pairs separated by less than 3 magnitudes in brightness) and place a lower limit on the fraction of dwarf galaxies ($10^{8}-10^{9.5}~M_{\odot}$) in such pairs of $\sim2\%$.    

We demonstrate that the low-redshift $z\sim 0.1$ pair fraction is $\sim 50\%$ higher than indicated by previous studies.  This result is due mainly to higher spectroscopic completeness and higher quality photometry from UKIDSS.  

We combine our low-redshift sample with a stellar-mass-selected sample from the RCS1 to measure the evolution of the pair fraction and merger rate for $L\sim L^*$ galaxies at $z<0.8$.  Assuming the pair fraction evolution may be parameterized as $f_{\rm{pair}} \propto (1+z)^m$, we find $m= 0.7\pm 0.1$, which constitutes a much flatter evolution than found in many other studies.  If the timescale for mergers is $\sim 0.3 - 0.5$~Gyr, and $50-100\%$ of pairs eventually merge, this implies the typical $L^*$ galaxy has undergone $\sim 0.2-0.8$ major mergers since $z=1$.  
 
We study the sample variance systematics in the pair fraction measurement in detail and conclude that apparent discrepancies between the different low-redshift samples in this study can be attributed to a combination of statistical errors and cosmic variance, and that the UKIDSS sample should be the most robust estimate at $f_{\rm{pair}}(z\sim 0.1) \approx 2 \pm 0.2\%$. The combination of the UKIDSS measurement with our RCS1 pair fraction estimates, and recent results published by the GAMA survey team, provide for a robust characterization of the pair fraction evolution to $z=0.8$.

\acknowledgements{
We thank the anonymous referee for comments and suggestions that helped to significantly improve this manuscript.
 
This research made use of the ``K-corrections calculator'' service available at http://kcor.sai.msu.ru/.

This work is based in part on photometric data products from the UKIRT
Infrared Deep Sky Survey (UKIDSS) and the Two Micron All Sky Survey (2MASS).

This study has made use of spectroscopic data products from the Two Degree Field Galaxy Redshift Survey (2DFGRS), the Six Degree Field Galaxy Redshift Survey (6DFGRS), the Two Micron Redshift Survey (2MR), and the Galaxy and Mass Assembly Survey (GAMA).  

The results presented here are based in part on a reanalysis of the Red Sequence Cluster Survey (RCS1) data.  The northern sky RCS1 data were obtained using the CFH12K camera on the Canada-France-Hawaii Telescope (CFHT), which is operated by the National Research Council of Canada, the Institut National des Sciences de l'Univers of the Centre National de la Recherche Scientifique of France, and the University of Hawaii.  The southern sky RCS1 data were obtained using the Mosaic II camera on the Cerro Tololo Inter-American Observatory (CTIO) 4~m Blanco telescope.

This publication makes use of data products from the Sloan Digital Sky Survey (SDSS). Funding for the SDSS and SDSS-II has been provided by the Alfred P. Sloan Foundation, the Participating Institutions, the National Science Foundation, the U.S. Department of Energy, the National Aeronautics and Space Administration, the Japanese Monbukagakusho, the Max Planck Society, and the Higher Education Funding Council for England. The SDSS Web Site is http://www.sdss.org/. The SDSS is managed by the Astrophysical Research Consortium for the Participating Institutions. The Participating Institutions are the American Museum of Natural History, Astrophysical Institute Potsdam, University of Basel, University of Cambridge, Case Western Reserve University, University of Chicago, Drexel University, Fermilab, the Institute for Advanced Study, the Japan Participation Group, Johns Hopkins University, the Joint Institute for Nuclear Astrophysics, the Kavli Institute for Particle Astrophysics and Cosmology, the Korean Scientist Group, the Chinese Academy of Sciences (LAMOST), Los Alamos National Laboratory, the Max-Planck-Institute for Astronomy (MPIA), the Max-Planck-Institute for Astrophysics (MPA), New Mexico State University, Ohio State University, University of Pittsburgh, University of Portsmouth, Princeton University, the United States Naval Observatory, and the University of Washington.

This work has made use of NASA's Astrophysics Data System and the NASA/IPAC Extragalactic Database.}

\end{document}

%% file: Table1.tex
\begin{deluxetable*}{lccccc}
\tabletypesize{\tiny}
\tablewidth{0pt}
\tablecaption{\label{pftable}Pair Fraction Data From Figure~\ref{pf}}
\tablehead{ \textbf{Source} & log$_{10}(L_{K,\rm{bin}})$\tablenotemark{a} &     $<$log$_{10}(L_K)>$\tablenotemark{b}    & $<z>$\tablenotemark{c}   &  $N_{\rm{tot}}$\tablenotemark{d}   & $f_{\rm{pair}}$\tablenotemark{e}}
\startdata

\textbf{UKIDSS ($K<16.3,~z<0.1$)} & $8.0~<L_K<~9.0$ & 8.72 & 0.004 & 146 & $0.014\pm 0.010$ \\
~~~~~~~~---  & $9.0~<L_K<10.0$ & 9.78 & 0.013 & 606 & $0.020\pm 0.006$ \\ 
~~~~~~~~--- & $10.0<L_K<10.2$ & 10.14 & 0.023 & 984 & $0.017\pm 0.004$ \\
~~~~~~~~---  & $10.2<L_K<10.6$  &10.46 & 0.031 & 3156 & $0.021\pm 0.003$\\
~~~~~~~~---  & $10.6<L_K<10.8$ & 10.71 & 0.042 & 3418 & $0.021\pm 0.003$ \\
~~~~~~~~---  & $10.8<L_K<11.2$  & 11.03 & 0.064 & 12193 & $0.022\pm 0.001$\\
~~~~~~~~---  & $11.2<L_K<11.3$  & 11.24 & 0.077 & 2388 & $0.020\pm 0.003$ \\
~~~~~~~~---  & $11.3<L_K<11.7$  & 11.38 & 0.079 & 2245 & $0.018\pm 0.003$\\
\\
\textbf{UKIDSS ($K<16.3,~0.1<z<0.25$)} & $10.8<L_K<11.4$ & 11.34 & 0.110 & 1652 & $0.022\pm 0.004$ \\
~~~~~~~~---  & $11.4<L_K<11.6$ & 11.48 & 0.128 & 1974 & $0.022\pm 0.003$ \\ 
~~~~~~~~--- & $11.6<L_K<12.0$ & 11.70 & 0.161 & 1009 & $0.019\pm 0.004$ \\
\\
\textbf{GAMA ($K<17,~z<0.15$)} & $9.0~<L_K<10.7$ & 10.47 & 0.041 & 872 & $0.018\pm 0.005$ \\
~~~~~~~~---  & $10.7<L_K<10.8$ & 10.75 & 0.064 & 361 & $0.019\pm 0.007$ \\ 
~~~~~~~~--- & $10.8<L_K<11.2$ & 11.03 & 0.089 & 2120 & $0.026\pm 0.004$ \\
~~~~~~~~---  & $11.2<L_K<11.4$  &11.28 & 0.116 & 1012 & $0.028\pm 0.005$\\
~~~~~~~~---  & $11.4<L_K<12.0$ & 11.50 & 0.119 & 404 & $0.025\pm 0.008$ \\
\\
\textbf{GAMA ($K<17,~0.15<z<0.3$)} & $11.0<L_K<11.4$ & 11.35 & 0.163 & 289 & $0.045\pm 0.012$ \\
~~~~~~~~---  & $11.4<L_K<11.6$ & 11.47 & 0.185 & 564 & $0.046\pm 0.009$ \\ 
~~~~~~~~--- & $11.6<L_K<12.0$ & 11.74 & 0.241 & 612 & $0.036\pm 0.008$ \\
\\
\textbf{2MASS ($K<15.36,~z<0.1$)} & $8.0~<L_K<~9.5$ & 9.20 & 0.004 & 192 & $0.019\pm 0.010$ \\
~~~~~~~~---  & $9.5~<L_K<10.4$ & 10.16 & 0.014 & 1024 & $0.021\pm 0.004$ \\ 
~~~~~~~~--- & $10.4<L_K<10.8$  & 10.64 & 0.026& 4864 & $0.019\pm 0.002$ \\
~~~~~~~~---  & $10.8<L_K<11.2$ &11.02  & 0.039 & 8985 & $0.021\pm 0.002$\\
~~~~~~~~---  & $11.2<L_K<11.5$  & 11.34 & 0.058& 5493 & $0.016\pm 0.002$ \\
~~~~~~~~---  & $11.5<L_K<12.0$  & 11.54 & 0.073 & 882 & $0.020\pm 0.005$\\
\\
\textbf{2MASS ($K<13.6,~z<0.06$)} & $9.3~<L_K<10.8$ & 10.55 & 0.010 & 1866 & $0.014\pm 0.003$ \\
~~~~~~~~--- & $10.8<L_K<11.2$ & 11.01& 0.017  & 3388 & $0.014\pm 0.002$ \\ 
 ~~~~~~~~--- & $11.2<L_K<11.4$ & 11.26 & 0.023 & 1311 & $0.014\pm 0.003$ \\
~~~~~~~~--- & $11.4<L_K<12.0$ &11.52 & 0.031  & 1800 & $0.009\pm 0.002$\\
\\

\enddata
\tablenotetext{a}{Bin luminosity range.}
\tablenotetext{b}{Mean luminosity of all galaxies in bin.}
\tablenotetext{c}{Mean redshift of all galaxies in bin.}
\tablenotetext{d}{Total number of candidate primary galaxies (see equation 1).}
\tablenotetext{e}{See equation 1.}

\end{deluxetable*}

%% file: Table2.tex
\begin{deluxetable*}{lccccc}
\tabletypesize{\scriptsize}
\tablewidth{0pt}
\tablecaption{\label{pfvztable}Pair Fraction Evolution Data From (Figure~\ref{pfvz})}
\tablehead{ \textbf{Source} & $M^*$ or $L_{\rm{K}}$ Selection & Redshift Range & $<z>$  &$N_{\rm{tot}}$\tablenotemark{a} & $f_{\rm{pair}}$\tablenotemark{b}}
\startdata

\textbf{UKIDSS ($K<16.3$)} &  $10.8 < \rm{Log_{10}}$$(L_K) < 11.2$ & $z<0.10$ & 0.064 & 12193 & $0.022\pm 0.003$ \\
\\
\textbf{GAMA ($K<17.0$)} &  $10.8 < \rm{Log_{10}}$$(L_K) < 11.2$ & $z<0.15$ & 0.089 & 2120 & $0.027\pm 0.012$ \\
\\
\textbf{2MASS ($K<15.36$)} &  $10.8 < \rm{Log_{10}}$$(L_K) < 11.2$ & $z<0.10$ & 0.039 & 8985 &$0.021\pm 0.005$ \\
\\
\textbf{2MASS ($K<13.6$)} &  $10.8 < \rm{Log_{10}}$$(L_K) < 11.2$ & $z<0.06$ & 0.018 & 3388 &$0.013\pm 0.006$ \\
\\
\textbf{RCS1} &  $10.87 < \rm{Log_{10}}$$(M^*) < 11.27$ & $0.25 < z < 0.48$& 0.38 & 11471 & $0.031\pm 0.005$\\
~~~~~~~~---  &  $10.75 < \rm{Log_{10}}$$(M^*) < 11.15$ & $0.48< z < 0.61$& 0.55 & 10825 & $0.029\pm 0.005$\\
~~~~~~~~--- &  $10.55 < \rm{Log_{10}}$$(M^*) < 10.95$ & $0.61 < z < 071$& 0.66 & 10733 & $0.030\pm 0.005$\\
~~~~~~~~---  &  $10.54 < \rm{Log_{10}}$$(M^*) < 10.94$ & $0.71 < z < 0.80$& 0.76 & 9794 & $0.036\pm 0.006$\\
\\
\textbf{KPAIR \citet{Xu12}} &  $10.6 < \rm{Log_{10}}$$(M^*_K) < 11$ & $z<0.1$ & 0.040 & 5826 & $0.014\pm 0.002$ \\
\\
\textbf{CPAIR \citet{Xu12}} &  $10.6 < \rm{Log_{10}}$$(M^*) < 11$ & $0.2 < z < 0.4$& 0.30 & 706 & $0.035\pm 0.011$\\
~~~~~~~~---  &  $10.6 < \rm{Log_{10}}$$(M^*) < 11$ & $0.4< z < 0.6$& 0.50& 955 & $0.025\pm 0.007$\\
~~~~~~~~--- &  $10.6 < \rm{Log_{10}}$$(M^*) < 11$ & $0.6 < z < 0.8$& 0.70& 1913 & $0.044\pm 0.008$\\
~~~~~~~~---  &  $10.6 < \rm{Log_{10}}$$(M^*) < 11$ & $0.8 < z < 1.0$& 0.90& 3229 &$0.055\pm 0.010$\\
\\
\textbf{GAMA-II~\citet{Robo14}\tablenotemark{c} } &  $10.4 < \rm{Log_{10}}$$(M^*) < 10.9$ & $0.05 < z < 0.1$& 0.08 & 1996 & $0.0213\pm 0.0036$\\
~~~~~~~~---  &  $10.4 < \rm{Log_{10}}$$(M^*) < 10.9$ & $0.1< z < 0.15$& 0.128& 6435 & $0.0230\pm 0.0021$\\
~~~~~~~~--- &  $10.4 < \rm{Log_{10}}$$(M^*) < 10.9$ & $0.15 < z < 0.2$& 0.177& 7398 & $0.0240\pm 0.0021$\\
\enddata
\tablenotetext{a}{Total number of candidate primary galaxies (see equation 1).}
\tablenotetext{b}{Errors include Poisson counting errors plus an estimate of cosmic variance (see Section~\ref{cosvar}).}
\tablenotetext{c}{Pair fraction data given here for GAMA-II represent those reported by \citet{Robo14} multiplied by a factor of $0.925$ to adjust from the 1:3 mass ratio used in GAMA-II to the 1:$10^{0.4}$ used here and for the fact that no minimum projected separation criterion was used for the GAMA-II sample. These corrections are explained in Section~\ref{mergerrate}.}
\end{deluxetable*}